\documentclass{jfm}

\usepackage{graphicx}
\usepackage{newtxtext}
\usepackage{newtxmath}
\usepackage{natbib}
\usepackage[makeroom]{cancel}
\usepackage{mathtools} 
\usepackage[normalem]{ulem} 
\usepackage{hyperref}
\hypersetup{
    colorlinks = true,
    urlcolor   = blue,
    citecolor  = black,
}
\usepackage{booktabs} 

\newcommand{\RomanNumeralCaps}[1]
\linenumbers

\def\la{\langle}
\def\ra{\rangle}
\def\bx{\boldsymbol{x}}
\def\bk{\boldsymbol{k}}
\def\bnabla{\boldsymbol{\nabla}}

\def\la{\left<}
\def\ra{\right>}

\def\U{{\cal U}}

\title{Statistics of near-inertial waves over a background flow via quantum and statistical mechanics}

\author{Alexandre Tlili\aff{1} \&
  Basile Gallet\aff{1} \corresp{\email{basile.gallet@cea.fr}}}

\affiliation{\aff{1}Université Paris-Saclay, CNRS, CEA, Service de Physique de l’Etat Condensé, 91191 Gif-sur-Yvette, France.}

\begin{document}
\maketitle

\begin{abstract}
We revisit the interaction of an initially uniform near-inertial wave (NIW) field with a steady background flow, with the goal of predicting the subsequent organization of the wave field. To wit, we introduce an exact analogy between the Young Ben Jelloul (YBJ) equation and the quantum dynamics of a charged particle in a steady electromagnetic field, whose potentials are expressed in terms of the background flow. We derive the time-averaged spatial distributions of wave kinetic energy, potential energy and Stokes drift in two asymptotic limits. In the `strongly quantum' limit where the background flow is weak compared to wave dispersion, we compute the wave statistics by extending a strong-dispersion expansion initially introduced by YBJ. In the `quasi-classical' limit where the background flow is strong compared to wave dispersion, we compute the wave statistics by leveraging the equilibrium statistical mechanics of classical systems. We compare our predictions to numerical simulations of the YBJ equation, using an instantaneous snapshot from a two-dimensional turbulent flow as the steady background flow. The agreement is very good in both limits. In particular, we quantitatively describe the preferential concentration of NIW energy in anticyclones. We predict weak NIW concentration in both asymptotic limits of weak and strong background flow, and maximal anticyclonic concentration for background flows of intermediate strength, providing theoretical underpinning to observations reported by Danioux, Vanneste and B\"uhler (Journal of Fluid Mechanics, 773, 2015).
\end{abstract}

\begin{keywords}

\end{keywords}

\section{Introduction\label{sec:intro}}

Atmospheric storms deposit momentum in the upper Ocean over a fast timescale and an extended spatial scale, as compared to the typical time and length scales of the balanced Ocean flow~\citep{d1995upper}. The initial Ocean response to such impulsive large-scale forcing consists of near-inertial waves (NIW) superposed to the pre-existing smaller-scale background geostrophic flow~\citep{d1985energy,alford2016near}. Over the following weeks the NIW energy gets redistributed as a result of wave propagation and dispersion, together with advection and refraction by the background flow~\citep{kunze1985near}. In the vertical direction, NIW are transferred to deeper regions~\citep{kunze1985near,Balmforth1998,asselin2020penetration}, potentially inducing mixing at the base of the mixed layer and in the deep ocean~\citep{Munk1998}. In the lateral directions, the background geostrophic flow rapidly imprints its spatial scale onto the NIW field, which acquires horizontal structure~\citep{Balmforth1998,conn2024interpreting}. The two processes are intimately connected, as horizontal structure in the wave-field speeds up the downward propagation of NIW energy~\citep{YBJ1997}.

This wave-mean flow interaction problem is challenging because the NIW field and background flow have comparable length-scales (at least during the initial stage of the evolution), which rules out the applicability of asymptotic methods based on spatial scale separation. In a breakthrough paper, \citet{YBJ1997} (YBJ in the following) leveraged the timescale separation instead, the inertial frequency being much faster than the inverse eddy-turnover time of the background flow. Through a multiple-timescale expansion they derived a reduced evolution equation -- now referred to as the YBJ equation -- governing the complex amplitude of the NIW field. The equation includes the contributions from advection and refraction by the background flow, together with wave dispersion. 

With this reduced framework at hand, various research questions have been investigated over the last decades, regarding both the one-way coupling between the waves and the background flow~\citep{Balmforth1998,llewellyn1999near,Danioux2015,danioux2016near,thomas2017near,asselin2020refraction,conn2025regimes} and the two-way coupling between the waves and the mean flow~\citep{xie2015generalised,wagner2016three,rocha2018stimulated,xie2020downscale,thomas2020turbulent,thomas2021forward,asselin2020penetration}. Restricting attention to one-way coupling, a particularly convenient idealized setup consists in studying solutions to the YBJ equation in a horizontally periodic domain, using a horizontally invariant initial condition for the wave field that mimics impulsive forcing by an atmospheric storm. We adopt this setup in the present study, with the goal of characterizing the subsequent organization of the NIW field over a steady background flow. Most previous studies have focused on the NIW kinetic energy, which constitutes the dominant contribution to the mechanical energy of the waves. The vast majority of the literature reports an accumulation of NIW energy in anticyclones, as initially inferred by~\citet{kunze1985near} using ray-tracing arguments, before being characterized through numerical studies of increasing complexity~\citep{lee1998inertial,asselin2020penetration,thomas2021forward,chen2021interaction,raja2022near} and observational data~\citep{jaimes2010near}. Theoretical insight regarding such accumulation was provided by \citet{Danioux2015} (DVB in the following), who identified a previously overlooked invariant of the system. Based on the conservation of this invariant, DVB argue that NIW kinetic energy should indeed accumulate in anticyclones. Surprisingly, however, their numerical simulations of the YBJ equation indicate that such accumulation of NIW in anticyclones is perhaps not as generic as initially thought. Indeed, they report clear accumulation of NIW energy in anticyclonic regions for background flows of intermediate speed only, while such accumulation was hardly noticeable for both fast and slow background flows. Why such an accumulation of NIW in anticyclones arises only in an intermediate, non-asymptotic range of flow speeds remains a puzzle that motivates the present study. More generally, we focus on the following questions:

\begin{enumerate}
\item What is the spatial distribution of NIW kinetic energy in the equilibrated state? Does NIW kinetic energy accumulate in anticyclonic regions, and how strong is this accumulation? 
\item What is the distribution of NIW potential energy? While subdominant as compared to NIW kinetic energy, the potential energy is directly related to the horizontal gradients of the wave field, and therefore to its ability  to propagate downwards in a 3D model.
\item What is the spatial distribution of the Stokes drift associated with the NIW wave field? 
\end{enumerate}

For weak background flows, we address these questions through a `strong-dispersion' asymptotic expansion. This expansion was initially introduced by YBJ, who computed the spatial distribution of NIW kinetic energy. We extend their pioneering work by computing the spatial distributions of NIW potential energy and Stokes drift, and by considering next-order corrections (Appendix \ref{app:strong_disperion_2nd_order_intermediate}).
For strong background flows, instead, our approach heavily relies on an exact analogy between the YBJ equation and the quantum dynamics of a charged particle in an inhomogeneous electromagnetic field. That the YBJ equation has the form of a Schr\"odinger equation was noticed early on by various authors, some of whom then adapted methods from quantum mechanics to compute some oscillatory eigenmodes of the YBJ equation (see e.g. the recent study by \citet{conn2025regimes}). Only partial interpretation of the Hamiltonian entering the analogous Schr\"odinger equation is provided in the literature, however. We fill this gap in section~\ref{sec:quantum_analogy} by insisting that the YBJ equation is rigorously analogous to the quantum dynamics of a charged particle in a steady electromagnetic field, whose scalar and vector potentials are expressed directly in terms of the streamfunction of the background flow. The closest analogy was made by \cite{Balmforth1998}, who clearly identified the analogous magnetic-field term. However, \cite{Balmforth1998} subsequently deemed the remaining potential term unphysical, making no further application of the analogy. Yet, the analogy proves particularly insightful in the limit of fast background flows, which corresponds to the classical limit of the quantum mechanics problem. Answering questions (i) through (iii) above reduces to determining the equilibrium statistics of a set of charged particles in inhomogeneous scalar and vector potentials, a task that we carry out using equilibrium statistical mechanics in section~\ref{sec:classical}.

\section{Near-Inertial Waves over a steady background flow}

\begin{figure}
    \centerline{\includegraphics[width=7cm]{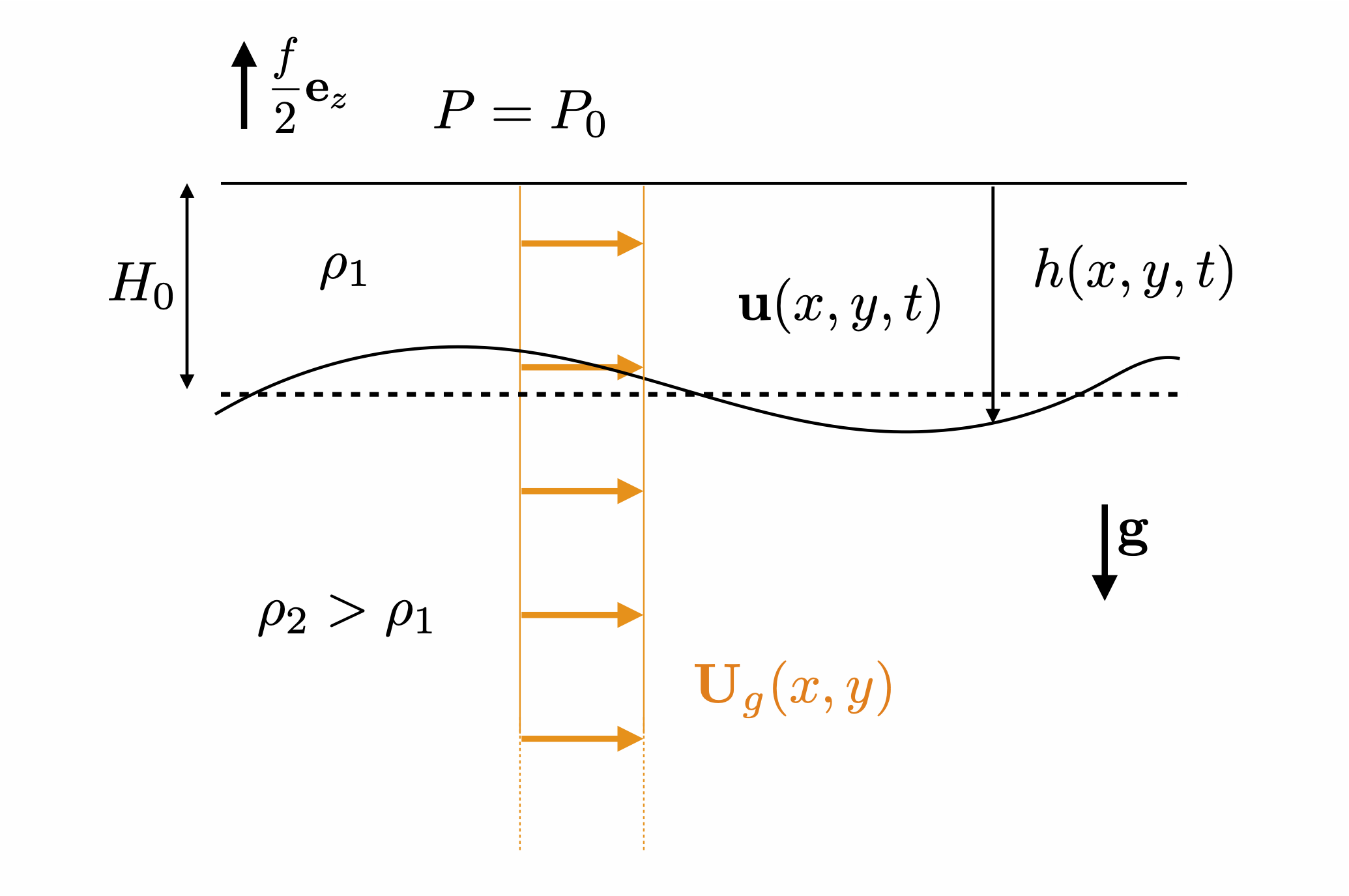}} 
   \caption{A two-layer model with an infinitely deep lower layer. The base state consists of a vertically invariant steady horizontal flow ${\bf U}_g(x,y)$ spanning both layers, together with a flat interface   between the two layers. We consider perturbations ${\bf u}(x,y,t)$ to the horizontal velocity in the upper layer only, whose depth is then denoted as $h(x,y,t)$. In line with the rigid-lid approximation, we neglect the fluctuations of the free surface as compared to $h$. \label{fig:schematicDVB}}
\end{figure}

\subsection{Wave dynamics in a shallow upper layer}

Consider the setup sketched in figure~\ref{fig:schematicDVB}, namely a two-layer model with upper-layer density $\rho_1$ and lower-layer density $\rho_2>\rho_1$, in a frame rotating at a positive rate $f/2$ around the vertical axis. Denoting time as $\tau$, the upper layer has depth $h(x,y,\tau)$, with $h=H_0$ in the rest state. The lower layer is infinitely deep. The base state consists of a steady, vertically invariant background flow ${\bf U}_g(x,y)$ spanning both layers. The background flow is in geostrophic balance with a vertically invariant lateral pressure gradient. It stems from a streamfunction $\psi(x,y)$, that is, ${\bf U}_g=[U_g(x,y),V_g(x,y),0]=-\bnabla \times (\psi \, {\bf e}_z)$. There is no deformation of the interface associated with such a vertically invariant balanced flow.
Following DVB, we consider the rotating shallow-water equations in the upper layer, linearized around the background balanced flow:
\begin{align}
\partial_\tau u +J(\psi,u) +u \partial_x U_g + v \partial_y U_g - f v & = -g' \partial_x h \, , \label{eq:udim}\\
\partial_\tau v + J(\psi,v) + u \partial_x V_g + v \partial_y V_g + f u & = -g' \partial_y h \, ,  \\ 
\partial_\tau h + J(\psi,h) + H_0 \bnabla \cdot {\bf u} & = 0 \, , \label{eq:hdim}
\end{align}
where ${\bf u}(x,y,\tau)=(u,v)$ denotes the horizontal velocity in the upper layer, $\bnabla=(\partial_x,\partial_y)$, the Jacobian operator is $J(s,q)=(\partial_x s) (\partial_y q) - (\partial_x q) (\partial_y s)$ and the reduced gravity is $g'=g\,(\rho_2-\rho_1)/\rho_1$ with $g$ the acceleration of gravity. 

In the absence of background flow, $U_g=V_g=0$, equations~(\ref{eq:udim}-\ref{eq:hdim}) support interfacial waves of frequency $\omega=f \sqrt{1 + k^2 \lambda^2}$, where $k$ denotes the wavenumber and $\lambda=\sqrt{g' H_0}/f$ is the small Rossby deformation radius associated with the shallow mixed layer. NIWs correspond to interfacial waves with wavelength much greater than $\lambda$, their frequency $\omega \simeq f (1 + k^2 \lambda^2/2)$ being close to $f$. The large-scale atmospheric forcing induces NIW in the upper ocean, and these waves remain near-inertial because the background geostrophic flow has a typical scale $L_\psi \gg \lambda$.

We non-dimensionalize equations~(\ref{eq:udim}-\ref{eq:hdim}) in such a way that the dimensionless fields and variables are ${\cal O}(1)$ at leading order in the expansion to come. Time is non-dimensionalized with $1/f$ and horizontal scales with $L_\psi$. The background-flow streamfunction is non-dimensionalized using its root-mean-square (rms) value $\psi_{\text{rms}}$, where the mean is performed over space. Denoting as $U_w$ the infinitesimal velocity-scale of the wave field, the wavy displacement of the interface scales as $H_0 U_w/(f L_\psi)$. With such scalings the dimensionless fields and variables read:
\begin{align}
    & \tau=\frac{\tilde{\tau}}{f} \, , \quad  \bx = L_{\psi} \tilde{\bx} \, , \quad (u,v)=U_w (\tilde{u},\tilde{v}) \, , \quad h = H_0\left( 1 + \frac{U_w}{f L_{\psi}}\tilde{h} \right) \, , \label{eq:scales} \\
    & \psi=\psi_\text{rms} \chi \, , \quad {\bf U}_g=\frac{\psi_\text{rms}}{L_\psi} \tilde{{\bf U}} \, \quad \text{with}  \quad \tilde{{\bf U}}=-\tilde{\bnabla} \times (\chi {\bf e}_z) \, , 
\end{align}
where tildes denote dimensionless quantities and derivatives with respect to dimensionless variables, and $\chi(x,y)$ is the dimensionless streamfunction. Denoting space average with angular brackets, the latter satisfies $\la \chi^2 \ra=1$. Substituting~(\ref{eq:scales}) into~(\ref{eq:udim}-\ref{eq:hdim}) leads to the dimensionless equations:
\begin{align}
    \partial_{\tilde{\tau}} \tilde{u} +Ro_\psi [\tilde{J}(\chi,\tilde{u})  -\tilde{u} \chi_{\tilde{x}\tilde{y}} - \tilde{v} \chi_{\tilde{y}\tilde{y}}] -  \tilde{v} & = - \epsilon \partial_{\tilde{x}} \tilde{h} \, ,  \label{eq:uadim}\\
    \partial_{\tilde{\tau}} \tilde{v} + Ro_\psi [\tilde{J}(\chi,\tilde{v}) + \tilde{u} \chi_{\tilde{x}\tilde{x}} + \tilde{v} \chi_{\tilde{x}\tilde{y}}] + \tilde{u} & = - \epsilon \partial_{\tilde{y}} \tilde{h} \, , \label{eq:vadim} \\ 
    \partial_{\tilde{\tau}} \tilde{h} + Ro_\psi \tilde{J}(\chi,\tilde{h}) +  \boldsymbol{\tilde{\nabla}} \cdot {\bf \tilde{u}} & = 0 \, ,
\label{eq:hadim}
\end{align}
where $\epsilon=(\lambda/L_\psi)^2$ denotes the Burger number and $Ro_\psi=\psi_\text{rms}/(f L_\psi^2)$ denotes the Rossby number of the background flow.

\subsection{The Young-Ben Jelloul (YBJ) equation}

Young \& Ben Jelloul (YBJ) consider the distinguished  asymptotic regime $\epsilon \ll 1$, $Ro_\psi \ll 1$, keeping a finite ratio $\gamma=Ro_\psi/\epsilon={\cal O}(\epsilon^0)$.
Through an asymptotic expansion recalled in Appendix~\ref{app:derivation}, YBJ  show that the horizontal velocity field $(u,v)$ consists of inertial oscillations whose complex amplitude slowly varies with time. They derive a reduced equation governing the modulation of this complex amplitude as a result of advection and refraction by the weak background flow, together with wave dispersion. Dropping the tildes to alleviate notations, for the present setup this procedure results in the following evolution equation for the \textit{demodulated} complex velocity field $M(\bx,t)=(u+iv)e^{i \tau}$:
\begin{equation}
   \partial_t M + 
    \underbrace{\vphantom{\frac{i}{2}} \gamma J(\chi, M)}_\text{advection}
    + \underbrace{\frac{i \gamma}{2}\left(\Delta \chi \right) M}_\text{refraction} 
    - \underbrace{\frac{i}{2}\Delta M}_{\mathclap{\text{dispersion}}}= 0 \, ,
    \label{eq:YBJ}
\end{equation}
where $\Delta = \partial_{xx} + \partial_{yy}$ is the Laplace operator and $t=\epsilon \tau$ is a slow time variable. Equation~(\ref{eq:YBJ}) is the simplest instance of the YBJ model. 
It involves the single dimensionless parameter $\gamma \geq 0$ characterizing the strength of the background flow relative to wave dispersion. In terms of dimensional variables, the expression of $\gamma$ is
\begin{equation}
    \gamma= \frac{\psi_\text{rms} f}{g' H_0} \, .
\end{equation}
Pure inertial oscillations with $M=\text{const.}$  are valid solutions to the YBJ equation (\ref{eq:YBJ}) in the absence of background flow only, that is for $\gamma=0$. For $\gamma \neq 0$ a uniform initial condition for $M$ evolves with time, developing some spatial structure as a result of refraction and advection by the background flow, and wave dispersion. We are interested in the fate of a uniform NIW field induced by a large-scale atmospheric storm. Equation~(\ref{eq:YBJ}) being linear and invariant to a uniform phase shift of the complex variable $M$, we focus on the initial condition $M(\bx,t=0)=1$ in the following.  

\section{The quantum analogy\label{sec:quantum_analogy}}

Early on, YBJ noticed the similarity between equation~(\ref{eq:YBJ}) and a Schr\"odinger equation, made more visible after multiplication by $i$:
\begin{equation}
  i  \partial_t M = - \frac{\Delta M}{2}  +\frac{\gamma}{2}(\Delta \chi) M -i \gamma J(\chi, M)   \, .
    \label{eq:YBJasSchro}
\end{equation}

\subsection{Particle in a steady electromagnetic field}

Consider a particle of mass $m$ and positive charge $q$ in a steady electromagnetic field whose potentials depend on $x$ and $y$. Using a set of units such that $m=q=\hbar=1$ (that is, using a non-dimensionalization based on $q$, $m$ and $\hbar$), the dimensionless Hamiltonian reads:
\begin{align}
H(\bx,{\bf p}) = \frac{1}{2}[{\bf p}-{\bf A}(x,y)]^2 +V(x,y) \, , \label{eq:classicalH}
\end{align}
where ${\bf A}(x,y)$ denotes the vector potential and $V(x,y)$ denotes the electrostatic potential, both dimensionless. The quantum dynamics of the particle are governed by the Schr\"odinger equation, $i \partial_t \phi = H\{ \phi \}$, where $\phi(x,y,t)$ denotes the wave function and the momentum ${\bf p}$ in the Hamiltonian~(\ref{eq:classicalH}) is replaced by the operator $-i \bnabla$. Now upon choosing dimensionless potentials that are related to the  streamfunction of the NIW problem through:
\begin{equation}
{\bf A}= \gamma \bnabla \times (\chi {\bf e}_z) \, \qquad \text{and} \qquad V=\frac{1}{2} \left(\gamma {\Delta \chi}- \gamma^2 {|\bnabla \chi|^2} \right) \, , \label{eq:potentials}
\end{equation}
the Schr\"odinger equation for the wave function $\phi(x,y,t)$ reduces precisely to the YBJ equation~(\ref{eq:YBJasSchro}) for the demodulated velocity $M(x,y,t)$. 
We conclude that there is an exact analogy between the YBJ equation and the quantum dynamics of a charged particle in the electromagnetic field given by the potentials~(\ref{eq:potentials}), the demodulated velocity $M(x,y,t)$ playing the role of the wave function of the charged particle. 
Within this analogy, the vector potential ${\bf A}$ equals minus the background flow velocity (and therefore ${\bf A}$ satisfies the Coulomb's gauge condition $\bnabla \cdot {\bf A}=0$) while the scalar potential $V$ equals one half the background flow vorticity, minus the background flow kinetic energy.
Table~\ref{tab:analogy} further lists analogous quantities between the two systems.

\begin{table}
\begin{center}
\begin{tabular}{|c|c|}
\hline
Quantum particle  & YBJ system \\ 
 \hline
wave function $\phi(x,y,t)$ &  $M(x,y,t)$  \\  
vector potential $\boldsymbol{A}(x,y)$ & $- \gamma {\bf U}(x,y)=\gamma \bnabla \times (\chi {\bf e}_z)$ \\
magnetic field $B(x,y){\bf e}_z=\bnabla \times {\bf A}$ & $- \gamma (\Delta \chi) {\bf e}_z$ \\
electric potential $V(x,y)$ & $\frac{1}{2} \left(\gamma \Delta \chi - \gamma^2 |\bnabla \chi|^2 \right)$ \\
conserved probability $\int_{\cal D} |\phi|^2 \mathrm{d}\bx$  & conserved wave action ${\cal A}=\langle |M|^2 \rangle$     \\
conserved energy $\langle \phi | H | \phi \rangle$  & conserved wave energy $E$, see. \eqref{eq:invariantE}   \\
\hline
\end{tabular}
\caption{Summary of the analogy between the Schr\"odinger equation for a charged particle (left) and the YBJ equation (right).\label{tab:analogy}}
\end{center}
\end{table}

\subsection{Conserved quantities}

There are two ways of determining the conserved quantities of the YBJ equation. One can directly deduce them from the equation, or one can readily infer them from the quantum analogy. Consider the YBJ equation inside a doubly periodic domain $(x,y)\in {\cal D}=[0,1]^2$. Multiplying the YBJ equation (\ref{eq:YBJ}) with $M^*$ before adding the complex conjugate and averaging over the domain ${\cal D}$ yields, after a few integrations by parts using the periodic boundary conditions:
\begin{align}
\frac{\mathrm{d} {\cal A}}{\mathrm{d}t}=0 \, ,  \qquad \text{with} \qquad {\cal A}=\la |M|^2 \ra \, , \label{eq:invariantA}
\end{align}
where the angular brackets denote space average over the domain ${\cal D}$.  Alternatively, the conservation of ${\cal A}$ is readily inferred from the quantum analogy, as ${\cal A}$ corresponds to the conserved total probability of finding the particle somewhere inside the domain ${\cal D}$. In the YBJ context, ${\cal A}$ corresponds to wave action, usually defined as the ratio of the wave energy to the wave frequency.
  The mechanical energy of NIWs is dominated by the kinetic energy $\la |M|^2 \ra$ (omitting the prefactor $1/2$), while the frequency is equal to $f$ to lowest order. The conservation of wave action thus reduces to the conservation of the space-averaged kinetic energy $\la |M|^2 \ra$ of the wave field.

The conservation of ${\cal A}$ is discussed in the original YBJ paper~\citep{YBJ1997}. Eighteen years later, a second independent conserved quantity was uncovered by DVB based on manipulations of the YBJ equation. Once again, this second invariant is readily inferred from the quantum analogy. Indeed, the Hamiltonian being time-independent, its expectation value is conserved over time: the mechanical energy of the charged particle is conserved. In the quantum context this expectation value is $\la \phi | H |\phi \ra$. In the YBJ context this quantity becomes $\int_{\cal D} M^* H\{ M \}\mathrm{d}\bx$, where $H\{ M \}$ is given by the rhs of~(\ref{eq:YBJasSchro}). The conserved quantity finally reads:
\begin{align}
\nonumber E & = \la M^* \left[ -\frac{1}{2} \Delta M  -i \gamma J(\chi,M) + \frac{\gamma}{2} (\Delta \chi)M \right]  \ra \\
& = \la 
\underbrace{\frac{|\bnabla M|^2}{2}}_{\text{potential}} 
+ \underbrace{\frac{\gamma \Delta \chi}{2} |M|^2 \vphantom{\frac{|\bnabla M|^2}{2}}}_{\text{refraction}} 
+ \underbrace{i \gamma \chi J(M^*,M) \vphantom{\frac{|\bnabla M|^2}{2}}}_{\text{advection}} \ra \ , \label{eq:invariantE}
\end{align}
where we have performed various integrations by parts using the periodic boundary conditions to obtain the second expression. We refer to (\ref{eq:invariantE}) as the wave energy. The various terms on the rhs of (\ref{eq:invariantE}) correspond to the potential energy, the contribution from the refractive term and the contribution from the advective term. In addition, the equation $\mathrm{d}E/\mathrm{d}t=0$ can be recast as an evolution equation for a single one of these energy terms, provided one substitutes the YBJ expression for $\partial_t M$ in the time derivative of the other forms of energy, see \cite{rocha2018stimulated}.

Strictly speaking, the total mechanical energy of the waves consists of a leading-order kinetic energy term, proportional to ${\cal A}$, and the weaker contributions gathered in $E$ above. In the present context ${\cal A}$ and $E$ are conserved independently. In the absence of background flow, $\gamma=0$, only the potential energy contribution $\la |\bnabla M|^2/2 \ra$ remains in~(\ref{eq:invariantE}).

\section{Organization of the NIW field over a steady background flow}

At this stage, one may reasonably object that we have made an analogy with a system that is perhaps less intuitive than the original system. We argue, however, that the analogy leads to various simple and useful observations. At the quantitative level, the analogy suggests methods to predict the wave statistics that will prove useful in section~\ref{sec:classical}.
At the qualitative level, the exact quantum analogy suggests a refinement of the arguments put forward by DVB. Indeed, focusing on the spatial distribution of wave action, DVB propose a partial quantum analogy: neglecting the advective term in~(\ref{eq:YBJasSchro}), the YBJ equation looks like a Schr\"odinger equation with a potential proportional to the vorticity $\gamma \Delta \chi$ of the background flow. DVB thus conclude that the particles will accumulate in the regions of lowest potential, which correspond to the anticyclones of the background flow. As mentioned in the introduction, this prediction is backed by {their numerical simulations for flows of intermediate strength only, whereas simulations with weak or strong background flows exhibit only a weak correlation between wave action and background flow vorticity. 

Such departures from the qualitative argument of DVB is to be expected from the exact quantum analogy in section~\ref{sec:quantum_analogy}. The full potential $V$ in~(\ref{eq:potentials}) consists of half the background flow vorticity, to which is added minus the flow kinetic energy. In the limit of fast background flow, the potential minima correspond to fast-flow regions, as opposed to anticyclones. If the particles were to accumulate in potential minima, they should end up in the regions of fastest background flow. However, it is also appropriate to question the underlying reasons for the accumulation of particles in potential minima. While a damped particle ends up in the potential well, a conservative particle accelerates as it reaches the potential minimum, spending very little time in the well.

With these questions in mind we revisit the spatial distribution of NIW over a background flow, based on theoretical predictions backed by numerical simulations.

\subsection{Numerical setup\label{sec:numerics}}

\begin{figure}
    \centering
    \includegraphics[width=\linewidth]{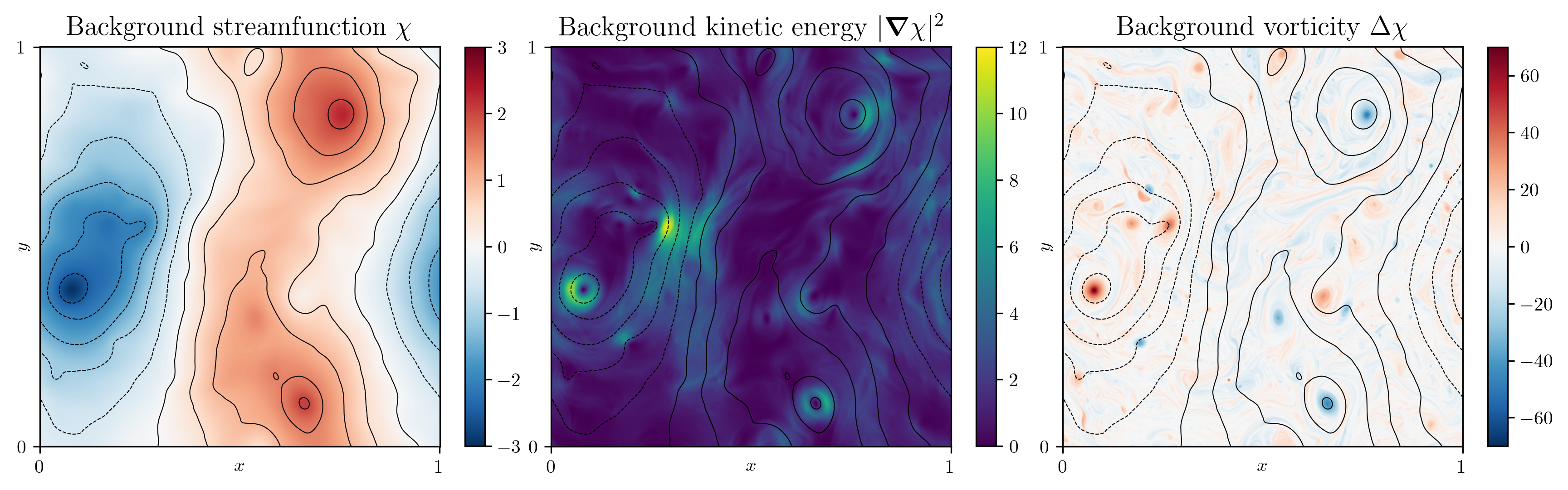}
    \caption{Steady background flow used in the numerical simulations of the YBJ equation: background streamfunction $\chi(x,y)$ (left), kinetic energy $|\bnabla \chi|^2$ (center) and vorticity field $\Delta\chi$ (right). The normalization is such that $\la \chi^2 \ra=1$ (see text). In all panels, the black contours correspond to streamlines of the background flow.}
    \label{fig:background_flow}
\end{figure}

The goal of the present study is to characterize the organization of the NIW field over a steady background flow, comparing the theoretical predictions to numerical simulations of the YBJ equation~(\ref{eq:YBJ}) in the doubly periodic domain ${\cal D}$. The simulations are performed using standard pseudo-spectral methods on a GPU with dealiasing and RK4 time-stepping. The timestep is fixed for a given simulation. No hyperviscosity is required, as the spatial spectrum of $M$ naturally exhibits an emergent cutoff wavenumber within the resolved scales. The  parameter values of all numerical simulations are reported in Appendix~\ref{app:numerics}. The initial condition is $M(\bx,t=0)=1$. For the steady background velocity field entering the equation, we use an instantaneous velocity field extracted from a simulation of the 2D Navier-Stokes equations, following the same forcing and dissipation protocols as described in~\citet{meunier2025effective}, albeit in a different parameter regime. 
We low-pass filter this frozen-in-time velocity field to remove excessively small scales with wavenumber $k \gtrsim 150$. We display the streamfunction, kinetic energy and vorticity of the resulting background flow in figure~\ref{fig:background_flow}. We stress the fact that this flow is time-independent in the YBJ equation.

\subsection{Quantities of interest}

In the following we mainly discuss the time-averaged spatial distributions of various forms of energy in the system. Denoting time-average with an overbar, we consider the spatial distribution of wave action $\overline{|M|^2}(\bf x)$, which also corresponds to the spatial distribution wave kinetic energy. We also consider the spatial distribution of 
the mean squared gradient of $M$, $\overline{|\bnabla M|^2}(\bf x)$. The latter being the dominant contribution to the NIW potential energy in both limits $\gamma \ll 1$ and $\gamma \gg 1$, we simply refer to it as the NIW potential energy in the following.
In the two-way coupled model derived in \cite{xie2015generalised}, $|\bnabla M|^2$ represents the NIW contribution to the total energy invariant, which makes it a quantity of interest for predictions. Together with these various forms of energy, we also discuss the time-averaged Stokes drift $\overline{{\bf u}_s}(\bx)$ induced by the wave field. The precise definition of the Stokes drift is deferred to Appendix~\ref{app:Stokes}, where we show that the time-averaged Stokes drift is related to the complex amplitude $M$ entering the YBJ equation through:

\begin{align}
\overline{{\bf u}_s}(\bx) = \frac{1}{4} \left[ i(\overline{M \bnabla M^*} - \overline{M^* \bnabla M}) + \bnabla \times (\overline{|M|^2} {\bf e}_z) \right]  \, . \label{eq:defus}
\end{align}
The dimensionless Stokes drift appearing in the equation above corresponds to the dimensional Stokes drift divided by $U_w^2/(f L_\psi)$.

\subsection{Two limiting regimes}

Once the flow structure $\chi(x,y)$ and the initial condition $M=1$ have been fixed, the only dimensionless parameter entering the problem is the strength $\gamma$ of the background flow. Guided by the quantum analogy, in the following we focus on two limiting situations of interest:
\begin{itemize}
\item $\gamma \ll 1$: this is the `quantum' or `strong-dispersion' limit. The background flow is weak and the dispersive effects in the YBJ equation~(\ref{eq:YBJ}) are strong.
\item $\gamma \gg 1$: this is the limit of `classical mechanics'. The YBJ equation is analogous to the dynamics of a quantum particle in the small-$\hbar$ limit. 
\end{itemize}

\section{The strong-dispersion `quantum' regime\label{sec:SDR}}

In the strong-dispersion limit $\gamma \ll 1$ the electrostatic potential reduces to:
\begin{align}
V(x,y)=- \gamma^2 \frac{|\bnabla \chi|^2}{2} +\frac{\gamma}{2} \Delta \chi  \simeq \frac{\gamma}{2} \Delta \chi \, .
\end{align}
In line with the intuition of DVB, the potential minima then correspond to the anticyclones of the background flow. Following YBJ we introduce the following low-$\gamma$ expansion for the NIW complex amplitude:
\begin{align}
M = {\cal M}(t) + \gamma \, m(x,y,t) + {\cal O}(\gamma^2) \, , \qquad \text{with }  \la m \ra = 0 \, . \label{eq:ansatzss}
\end{align}
In (\ref{eq:ansatzss}) the homogeneous initial condition has evolved into an ${\cal O}(1)$ homogeneous part ${\cal M}(t)$ of the solution, together with a weaker  mean-zero spatial modulation $\gamma \, m(x,y,t)$ induced by the weak background flow. Both ${\cal M}$ and $m$ are ${\cal O}(1)$ in the expansion above. Averaging the YBJ equation (\ref{eq:YBJ}) over space simply leads to $\partial_t {\cal M}= 0 +{\cal O}(\gamma^2)$: the spatially homogeneous part of the solution is time-independent to lowest order, and using the initial condition we obtain ${\cal M}=1$. To ${\cal O}(\gamma)$, the YBJ equation~(\ref{eq:YBJ}) then yields:
\begin{align}
\partial_t m  + {\frac{i}{2} \Delta \chi} {- \frac{i}{2} \Delta m}= 0 \, .
\label{eq:sdr_order1_eq}
\end{align}
The general time-dependent solution to~\eqref{eq:sdr_order1_eq} is
\begin{align}
m(x,y,t) = \chi(x,y) + \tilde{m}(x,y,t) \, ,
\end{align}
where the term $\tilde{m}(x,y,t)$ oscillates in time with vanishing time average. Introducing a Fourier decomposition of the background streamfunction as $\chi(x,y)=\sum_{\boldsymbol{k}} \hat{\chi}_{\boldsymbol{k}} e^{i\boldsymbol{k \cdot x}}$ and imposing that $m$ vanishes at $t=0$ (in line with the initial condition $M(\bx,t=0)=1$) gives
\begin{align}
\tilde{m}(x,y,t) = -\sum_{\boldsymbol{k}} \hat{\chi}_{\boldsymbol{k}} e^{i\boldsymbol{k \cdot x} - ik^2t/2}\ \, . \label{eq:exprmtilde}
\end{align}
$\tilde{m}$ above is neglected in the original derivation by YBJ, while being included by DVB.
The approximate solution for the complex demodulated velocity reads:
\begin{align}
M \simeq 1+\gamma \chi(x,y) + \gamma \tilde{m}\, , \label{eq:approxMSDR}
\end{align}
leading to the following approximate expression for the time-averaged distribution of wave action:
\begin{align}
\overline{|M|^2}(\bx) =  1+2 \gamma \chi(x,y) +{\cal O}(\gamma^2)\, . 
\label{eq:KESDR}
\end{align}
This perturbative computation of the distribution of NIW kinetic energy was initially obtained by YBJ. Equation~(\ref{eq:KESDR}) shows that, although the potential minima correspond to anticyclonic regions, the distribution of wave kinetic energy (or wave action) is modulated by the streamfunction of the flow, the regions of maximal wave kinetic energy corresponding to the regions of maximal streamfunction. In the particular case of a monoscale flow, where $\chi(x,y)$ is an eigenmode of the Laplace operator $\Delta$,  the vorticity is directly proportional to  $-\chi$: regions of strong $\chi$ indeed correspond to anticyclonic regions, confirming the intuition of DVB. For multiscale flows involving a broad range of scales, however, the streamfunction can differ very much from the vorticity field (see figure~\ref{fig:background_flow}).

We now extend the pioneering analysis of YBJ by computing additional quantities beyond the sole kinetic energy. As a first example, the time-averaged contribution from the potential energy to the energy invariant $E$ is given by (omitting the prefactor $1/2$):
\begin{align}
\overline{|\bnabla M|^2}(\bx) = \gamma^2 \left( |\bnabla \chi|^2 + \overline{|\bnabla \tilde{m}|^2} \right) \, , \label{eq:gradM2SDR}
\end{align}
to lowest order in $\gamma$. While the oscillatory part ${\tilde m}$ of the solution is irrelevant to compute the distribution of wave action~(\ref{eq:KESDR}), it arises at leading order in the time-averaged distribution of potential energy, see equation~(\ref{eq:gradM2SDR}).

As a second example, consider the time-averaged Stokes drift. After inserting the decomposition~(\ref{eq:approxMSDR}), equation~(\ref{eq:defus}) yields:
\begin{align}
\overline{{\bf u}_s}(\bx) = \frac{1}{4} \bnabla \times ( \overline{|M|^2} {\bf e}_z)  = - \frac{\gamma}{2} {\bf U}(x,y) + {\cal O}(\gamma^2)\, , \label{eq:usSDR}
\end{align}
where we have inserted expression~(\ref{eq:KESDR}) to obtain the last equality. We conclude that the time-averaged Stokes drift is opposite to the background flow (the dimensional Stokes drift, obtained by multiplying~(\ref{eq:usSDR}) with $U_w^2/(f L_\psi)$, is also quadratic in NIW velocity-scale $U_w$).

In figure~\ref{fig:comparison_strong_dispersion} we compare the predictions above for $\overline{ |M|^2}(\bx)$, $\overline{|\bnabla M|^2}(\bx)$ and $\overline{{\bf u}_s}(\bx)$ to the output of a numerical simulation of the YBJ equation~(\ref{eq:YBJ}) with a weak background flow, $\gamma=0.05$, following the setup described in section~\ref{sec:numerics}. The agreement between the predictions and the numerical results is excellent. This further confirms that 
the NIW kinetic energy $\overline{|M|^2}(\bx)$ develops some structure at the large scale of the background streamfunction $\chi$, rather than the scale of the background vorticity $\Delta \chi$.

\begin{figure}
    \centering
    \includegraphics[width=\linewidth]{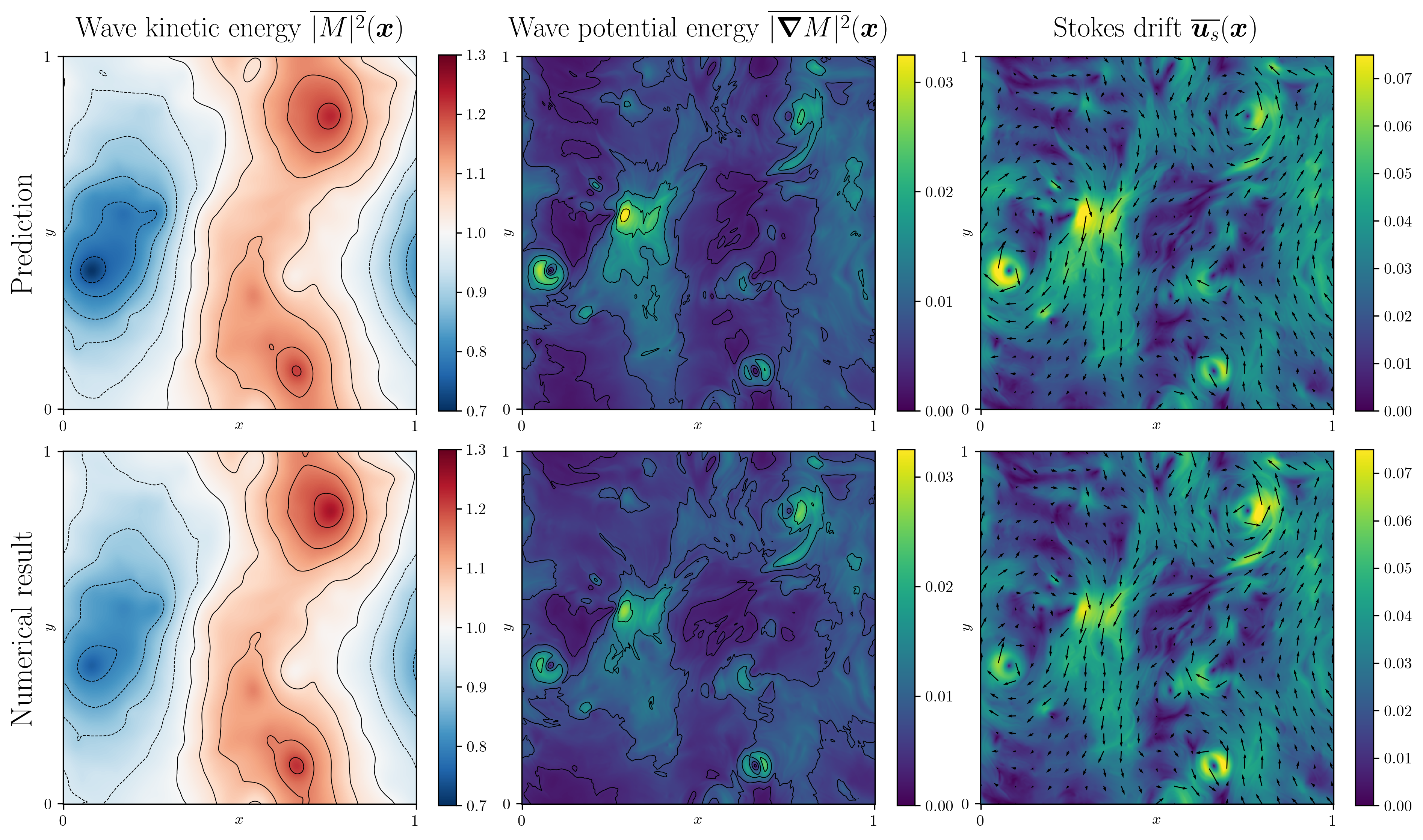}
    \caption{Time-averaged spatial distributions of NIW kinetic energy (left), potential energy (center) and Stokes drift (right). The top row corresponds to the predictions~(\ref{eq:KESDR}-\ref{eq:usSDR}) from the low-$\gamma$ asymptotic expansion. The bottom row corresponds to a numerical simulation in the strong-dispersion regime ($\gamma=0.05$). Isovalues are indicated with black contours with identical levels and colorbars for predictions and observations.}
    \label{fig:comparison_strong_dispersion}
\end{figure}

\section{The strong-advection `classical' regime\label{sec:classical}}

\begin{figure}
    \centerline{\includegraphics[width=6 cm]{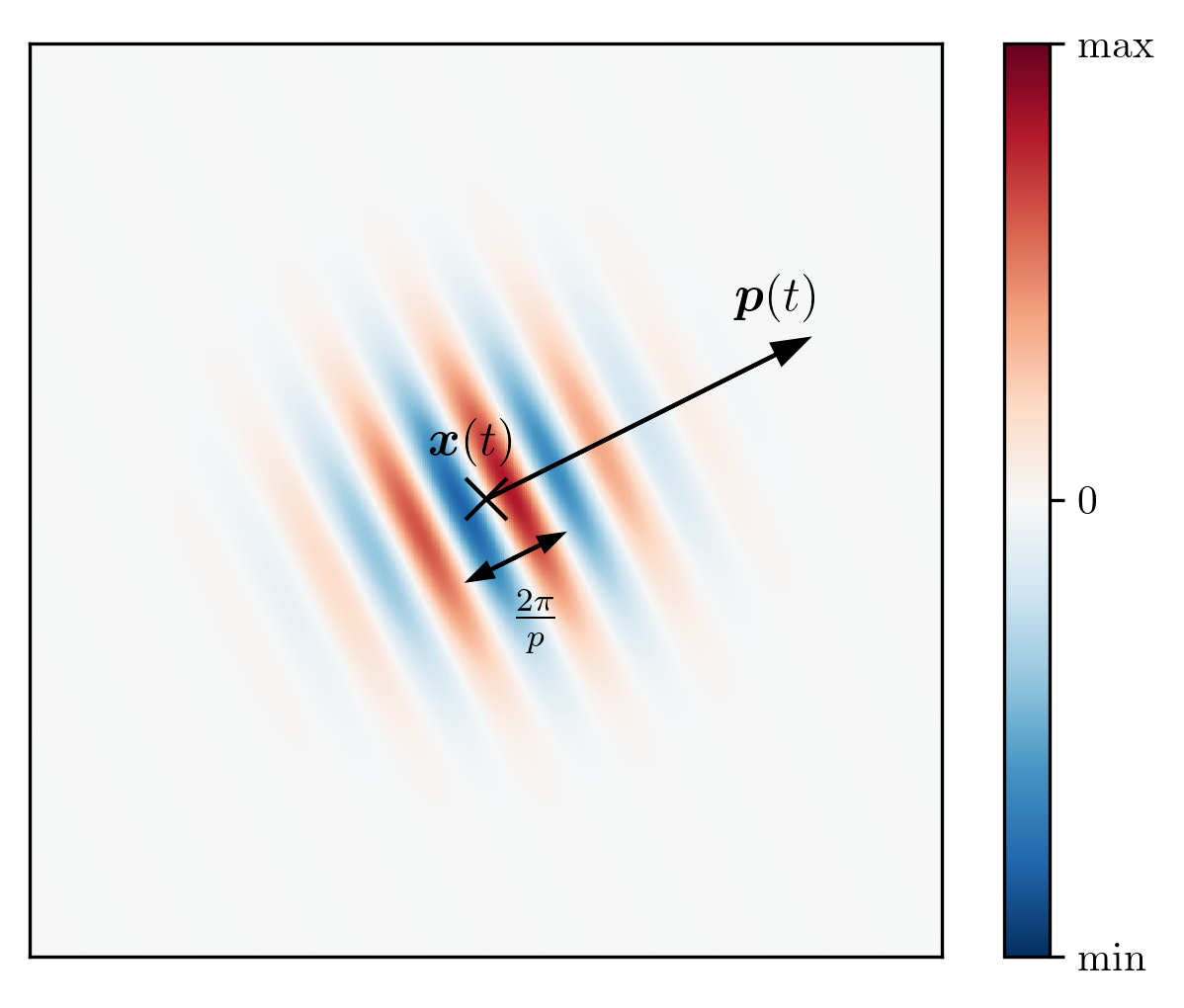}} 
   \caption{A narrow wave packet with mean position $\bx(t)$ and wavevector ${\bf p}(t)$ behaves like a charged classical particle in a steady 2D electromagnetic field. \label{fig:wavepacket}}
\end{figure}

Far fewer results are available in the strong-advection regime, $\gamma \gg 1$, in terms of predictions for the spatial distributions of NIW statistics. In this limit the potential reduces to:
\begin{align}
V(x,y)=- \frac{\gamma^2}{2} |\bnabla \chi|^2 +\frac{\gamma}{2} \Delta \chi  \simeq - \frac{\gamma^2}{2} |\bnabla \chi|^2 =-\frac{\gamma^2}{2} {\bf U}^2 \, . \label{eq:approxVcalssical}
\end{align}
The potential wells thus correspond to the local maxima of the kinetic energy of the background flow. $\gamma \gg 1$ is also the ray-tracing limit~\citep{buhler2014waves}, where the trajectories of compact wave packets are determined based on a WKB expansion. We readily infer the resulting ray-tracing equations from the quantum analogy: this is the limit of classical mechanics. A compact wave packet localized at $\bx(t)$ corresponds to a charged classical particle subject to a Lorentz force, and Newton's third law yields:
\begin{align}
m \ddot\bx= q ({{\bf E}+{\dot\bx}\times{\bf B}}) \, ,
\end{align}
where ${\bf E}=-\bnabla V$ denotes the electric field, ${\bf B}=\bnabla \times {\bf A}$ denotes the magnetic field (dimensional versions), and we have explicitly written the mass $m$ and the charge $q$ to highlight the analogy.

As mentioned above, it is far from obvious that the conservative dynamics of such classical particles should lead to accumulation in the potential wells. That is, one should not hastily conclude from (\ref{eq:approxVcalssical}) that the particles -- and thus the NIW kinetic energy -- will accumulate in the fast-flow regions. Instead, a better-suited framework to infer the statistics of such classical particles is the statistical mechanics of equilibrium systems.

\subsection{Ergodic theory and microcanonical ensemble}

Instead of Newton's third law, the statistical mechanics of equilibrium system starts from the classical version of the Hamiltonian~(\ref{eq:classicalH}). Hamilton's equations govern the evolution of a narrow wave packet located at $\bx(t)=[x(t),y(t)]$ with wavevector ${\bf p}(t)=[p_x(t),p_y(t)]$ (see sketch in figure~\ref{fig:wavepacket}):
\begin{align}
\frac{\mathrm{d} \bx}{\mathrm{d}t}= \frac{\partial H}{\partial {\bf p}} \, , \qquad \frac{\mathrm{d} {\bf p}}{\mathrm{d}t}= - \frac{\partial H}{\partial \bx} \, , \label{eq:Hequations}
\end{align}
where the equations are to be understood componentwise. Consider a cloud of initial conditions in the phase space $(x,y,p_x,p_y)$. Liouville's theorem states that, following the Hamiltonian evolution~(\ref{eq:Hequations}), the cloud will deform in phase space conserving its initial volume. In other words, the volume in phase space is conserved by the dynamics because equations~(\ref{eq:Hequations}) correspond to an incompressible flow in phase space. 

Consider now an ensemble of particles with the same initial energy $E_0$. Because energy is conserved, these particles only have access to the hypersurface $H(\bx,{\bf p})=E_0$ in phase space. 
Like a patch of dye getting homogenized by a chaotic flow and achieving uniform concentration in the long-time limit, we expect the Hamiltonian phase-space flow~(\ref{eq:Hequations}) to homogenize a cloud of initial conditions with initial energy $E_0$ over the hypersurface $H(\bx,{\bf p})=E_0$.
Introducing a probability density ${\cal P}(\bx,{\bf p})$ such that ${\cal P}(\bx,{\bf p})\mathrm{d}\bx \mathrm{d}{\bf p}$ is the probability for a particle to be in a phase-space volume $\mathrm{d}\bx \mathrm{d}{\bf p}$ around the point $(\bx,{\bf p})$, this ergodic assumption translates into:

\begin{align}
 {\cal P}(\bx,{\bf p}) = {\cal C} \, \delta[H(\bx,{\bf p})-E_0]\, , \label{eq:Probameasure}
\end{align}
where the constant ${\cal C}$ is a normalization factor. 
The validity of the ergodic prescription~(\ref{eq:Probameasure}) is a lively topic of research at the crossroads of mathematics and physics, known as `quantum chaos'~\citep{berry1977regular}. Rigorous mathematical results have been obtained based on the asymptotic behavior of the Wigner transform of the wavefunction in the classical limit~\citep{voros1976semi}. A detailed discussion of this topic goes beyond the scope of the present study. Instead, we motivate~(\ref{eq:Probameasure}) based on the microcanonical ensemble prescription of equilibrium statistical mechanics~\citep{bouchet2012statistical}, which is expected to correctly describe the statistics of the quantum system in the classical limit $\gamma \gg 1$. In the following, we thus assume that the ergodic assumption holds and we replace long-time averages with averages in phase space using the probability density~(\ref{eq:Probameasure}).

\subsection{Distribution of NIW kinetic energy}

As a first illustration, let us determine the time-averaged distribution of NIW kinetic energy (or wave action) $\overline{|M|^2}(\bx)$ using an average in phase space. According to table~\ref{tab:analogy}, $\overline{|M|^2}(\bx)$ corresponds to the time-averaged probability of finding the quantum particle at location $\bx$. And in the classical limit, this reduces to the probability of finding the classical particle at location $\bx$ regardless of its momentum ${\bf p}$. Using the microcanonical probability density~(\ref{eq:Probameasure}), the latter probability is given by:
\begin{align}
\overline{|M|^2}(\bx) & = \int_{\bx'\in {\cal D}  , \, {\bf p}\in \mathbb{R}^2} \underbrace{\delta(\bx'-\bx)}_{\text{particle located at $\bx$}} \, { {\cal P}}(\bx',{\bf p}) \, \mathrm{d}\bx' \mathrm{d}{\bf p} \\
& = {\cal C} \int_{{\bf p}\in \mathbb{R}^2}  \,\delta \left[ \frac{1}{2}({\bf p}+\gamma{\bf U}(\bx))^2  + V(\bx) -E_0  \right] \, \mathrm{d}{\bf p} \, .
\end{align}
Changing integration variable to ${\bf K}={\bf p}+\gamma{\bf U}(\bx)$ with norm $K=|{\bf K}|$, the integral becomes:
\begin{align}
\overline{|M|^2}(\bx) & = {\cal C} \int_{K \in \mathbb{R}^+}  \,\delta \left[ \frac{1}{2}K^2  + V(\bx) -E_0  \right] 2 \pi K \, \mathrm{d}K \\
 & = 2 \pi {\cal C} \int_0^\infty  \,\delta \left[ s  + V(\bx) -E_0  \right] \, \mathrm{d}s \, ,
\end{align}
where we changed integration variable again to $s=K^2/2$. The resulting integral equals one if $V(\bx) -E_0<0$ and zero if $V(\bx) -E_0>0$, that is:
\begin{align}
\overline{|M|^2}(\bx) & = 2 \pi {\cal C}  \, {\cal H}[E_0-V(\bx)]  \, , \label{eq:tempheavyside}
\end{align}
where ${\cal H}$ denotes the Heavyside function.

The initial energy of the particles is estimated by inserting the initial condition $M(x,y,t=0)=1$ into expression~(\ref{eq:invariantE}) for the energy. Only the term $\gamma (\Delta \chi)|M|^2/2$ remains: the local initial energy is of the order of the local vorticity and therefore it scales as $\gamma$. By contrast, the potential~(\ref{eq:approxVcalssical}) has much greater magnitude, of order $\gamma^2$, and it is negative almost everywhere. We conclude that the initial energy is negligible as compared to the potential $V<0$ in the limit  $\gamma \gg 1$ of interest here: $E_0 \simeq 0$ (see Appendix~\ref{app:KEdist} for a refined calculation taking into account the narrow distribution of $E_0$ around zero). To a good approximation, ${\cal H}[E_0-V(\bx)]=1$ almost everywhere, and we thus predict a uniform distribution of NIW kinetic energy, $\overline{|M|^2}(\bx) = 2 \pi {\cal C}$. Because of action conservation, the space average of $|M|^2$ is conserved and equal to one. We thus obtain ${\cal C}=1/(2 \pi)$, the prediction for the time-averaged spatial distribution of kinetic energy being simply:
\begin{align}
\overline{|M|^2}(\bx) & = 1 \, . \label{eq:predictionM2}
\end{align}
Somewhat surprisingly, based on statistical mechanics we predict a uniform distribution of NIW kinetic energy, despite the spatial structure of the potential~(\ref{eq:approxVcalssical}). This long-time behavior contrasts with the early-time behavior of the system, as described e.g. in DVB and in \citet{rocha2018stimulated}. At early time, the uniform initial condition for $M$ is affected predominantly by the refraction term, which induces strong phase gradients driving an action flux towards the center of anticyclones. For subsequent times, however, the advection term comes into play and, for $\gamma \gg 1$, DVB show that the dominant balance in the YBJ equation is eventually between advection and dispersion. Similarly, one can check that the refraction term plays a negligible role in the line of arguments leading to the uniform prediction~(\ref{eq:predictionM2}) from ergodic theory. In section~\ref{sec:refined}, we address the influence of the small refraction term in more details to refine the prediction~(\ref{eq:predictionM2}).

\subsection{Distribution of NIW potential energy}

As a second illustration of the statistical mechanics approach, we consider the time-averaged spatial distribution of NIW potential energy, $\overline{|\bnabla M|^2}(\bx)$. The dimensionless momentum operator being $-i \bnabla$ according to the quantum analogy, the NIW potential energy is analogous to the expectation value of the squared momentum. Alternatively, based on the sketch in figure~\ref{fig:wavepacket} one estimates $\bnabla M \simeq i {\bf p} M$ and $|\bnabla M|^2 \simeq p^2 |M|^2$, in line with the standard WKB approximation. We thus seek the averaged squared momentum of the particles located at $\bx$. In phase space this average reads:
\begin{align}
\overline{|\bnabla M|^2}(\bx) & = \int_{\bx'\in {\cal D}, \, {\bf p}\in \mathbb{R}^2} \underbrace{p^2}_{\text{squared momentum}} \delta(\bx'-\bx) \, {\cal P}(\bx',{\bf p}) \, \mathrm{d}\bx' \mathrm{d} {\bf p} \, , \label{eq:tempintPE} \\
& = 2 \gamma^2 {\bf U}(\bx)^2 \, , \label{eq:predictionPE}
\end{align}
the integration in phase space being detailed in Appendix~\ref{app:Integrals}. We conclude that the spatial distribution of NIW potential energy is given by the kinetic energy distribution of the background flow, with an accumulation of NIW potential energy in fast-flow regions. 

\subsection{Time-averaged Stokes drift}

As a last illustration of the statistical mechanics approach, we consider the time-averaged Stokes drift~(\ref{eq:defus}). In the $\gamma \gg 1$ limit, substitution of the prediction~(\ref{eq:predictionM2}) for $\overline{|M|^2}(\bx) $ into~(\ref{eq:defus}) shows that the second term vanishes. Inserting again the estimate $\bnabla M \simeq i {\bf p} M$, the time-averaged Stokes drift~(\ref{eq:defus}) reduces to:
\begin{align}
\overline{{\bf u}_s}(\bx) = \frac{i}{4} (\overline{M \bnabla M^*} - \overline{M^* \bnabla M}) \simeq \frac{1}{2} \overline{{\bf p} |M|^2}(\bx)  \, . \label{eq:ustemp}
\end{align}
Once again, we assume ergodicity to compute the average appearing on the rhs in phase space. The integration in phase space is deferred to Appendix~\ref{app:Integrals}, the end result being:
\begin{align}
\overline{{\bf u}_s}(\bx) =- \frac{\gamma}{2} {\bf U}(\bx)  \, . \label{eq:usSAR}
\end{align}
It is remarkable that we obtain the same prediction for the time-averaged Stokes drift in the strong-dispersion limit and in the strong-advection limit, see (\ref{eq:usSDR}) and (\ref{eq:usSAR}), although these two predictions arise from different terms in the expression~(\ref{eq:defus}) of the Stokes drift.

\subsection{Comparison to numerical simulations}

\begin{figure}
    \centering
    \includegraphics[width=0.9\textwidth]{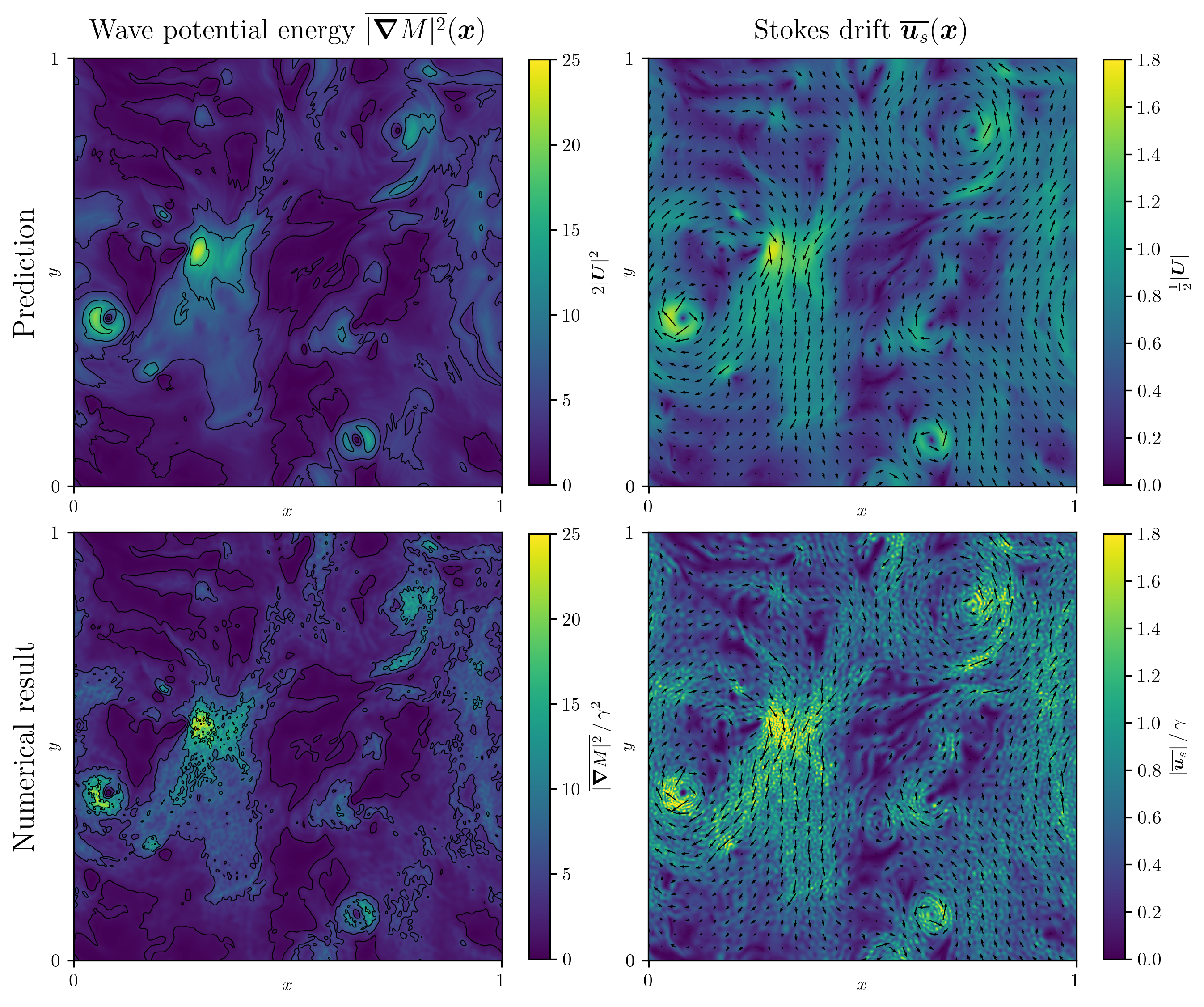}
    \caption{Top row: ergodic predictions for the time-averaged NIW potential energy $\overline{|\bnabla M|^2}(\bx)$ (left) and Stokes' drift $\overline{\boldsymbol{u}_s}(\bx)$ (right). Bottom row: same fields extracted from a numerical run with $\gamma=30$. In the left-hand column, black contours indicate isovalues 1, 4, 9 and 16.}
    \label{fig:comparison_usPE_SAR}
\end{figure}

\begin{figure}
    \centering
    \includegraphics[width=\textwidth]{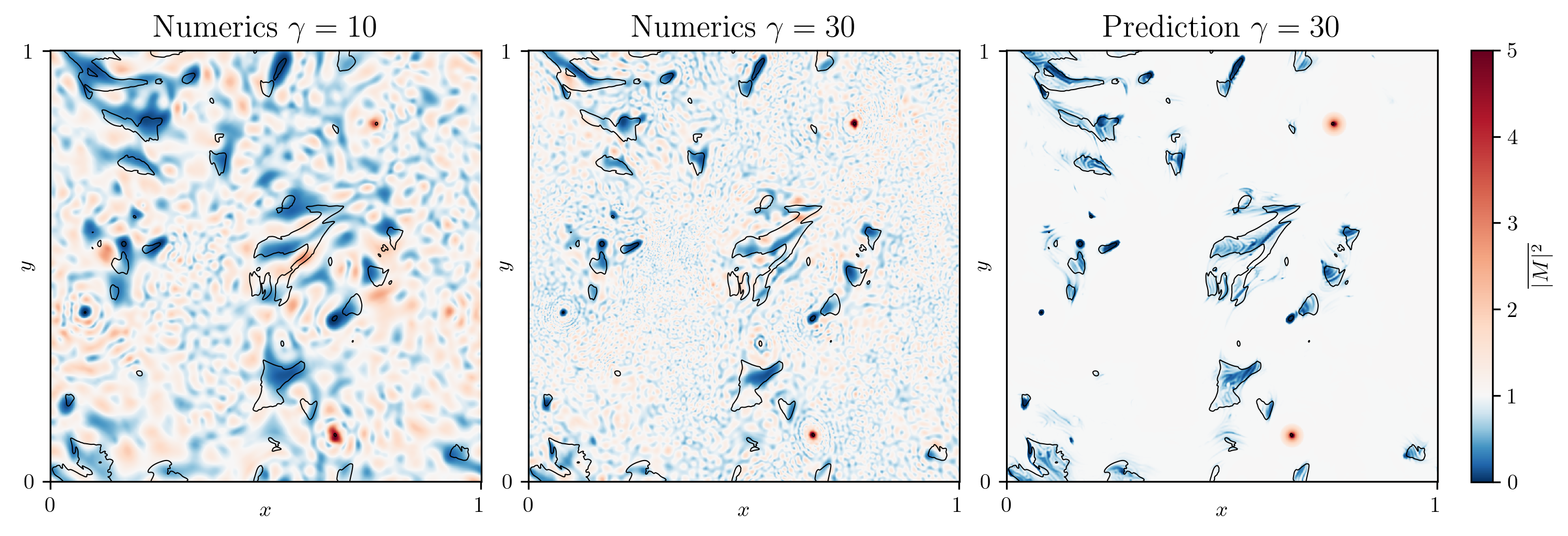}
    \caption{Left: Time-averaged NIW kinetic energy $\overline{|M|^2}$ from a numerical simulation with $\gamma=10$. Center : Idem for $\gamma = 30$. Right: Refined ergodic prediction~(\ref{eq:KEwave_SAR_fullpred}) for $\gamma = 30$, taking into account non-uniform initial energy and trapping in the two main anticyclones. The colorscale lightness varies linearly on $[0,1]$ and on $[1,5]$ to keep $\overline{|M|^2}=1$ in white. Black curves indicate the isovalue $1/3$ of the rms value of $|\mathbf{U}|$. These curves show that deviations from the uniform distribution preferentially occur in regions of slow background flow.}
    \label{fig:comparison_KE_SAR}
\end{figure}

In figure~\ref{fig:comparison_usPE_SAR},  we compare the time-averaged spatial distributions of potential energy and Stokes drift with the ergodic predictions~(\ref{eq:predictionPE}) and~(\ref{eq:usSAR}). The agreement is very good in both cases, both at the qualitative and at the quantitative level, showing that NIW packets indeed tend to behave in an ergodic fashion in the strong-advection limit. While the large-scale structure of both fields is accurately captured by the ergodic theory, one can notice some small-scale roughness or `granularity' in the numerical fields in figure~\ref{fig:comparison_usPE_SAR}. This phenomenon does not seem to disappear over longer time average. Rather, it stems from some form of interference pattern, reminiscent of the patterns obtained in studies of quantum chaos~\citep{voros1976semi,berry1977regular,nonnenmacher2013anatomy}. The latter studies suggest that the granularity arises at the de Broglie wavelength of the system and that classical statistics apply to quantities averaged over a few de Broglie wavelengths. Equations~(\ref{eq:ustemp}) and~(\ref{eq:usSAR}) point to the scaling $p\sim \gamma$ for the typical momentum of the particle, and therefore a de Broglie wavelength that scales as $1/p \sim 1/\gamma$ (remembering that $\hbar=1$ with our choice of units). 
The same estimate is readily obtained by DVB who show that, in the large-$\gamma$ regime, the characteristic scale of $M$ is set through a balance between advection and dispersion.
Some slight and slow time-dependence in the position of the vortices of the background flow, or an ensemble average over a family of slightly structured initial conditions for $M$, may be sufficient to disrupt the precise phase relations producing the interference pattern, thus smoothing out the observed granularity.

Consider now the spatial distribution of NIW kinetic energy, illustrated in figure~\ref{fig:comparison_KE_SAR}. The naive ergodic prediction~(\ref{eq:predictionM2}) corresponds to a uniform distribution of wave kinetic energy. There is reasonable agreement with this prediction: for large $\gamma$, the fields in figure~\ref{fig:comparison_KE_SAR} feature an extended uniform white region with $\overline{|M|^2}(\bx) \simeq 1$. However, locally we observe some strong departures from this uniform background. These departures are located near the points of vanishing kinetic energy of the background flow, as illustrated by the contour of low $|{\bf U}|$. More precisely, we distinguish between two types of regions:
\begin{itemize}
\item In regions of vanishing ${\bf U}^2(\bx)$ with cyclonic background vorticity $\gamma \Delta \chi (\bx)>0$, we observe a \textit{deficit} of NIW kinetic energy (blue regions).
\item In regions of vanishing ${\bf U}^2(\bx)$ with anticyclonic background vorticity $\gamma \Delta \chi (\bx)<0$, we observe an \textit{excess} of NIW kinetic energy, see the narrow red regions in the two strongest anticyclones in figure~\ref{fig:comparison_KE_SAR}.
\end{itemize}

\subsection{Refined ergodic prediction, and anticyclonic trapping\label{sec:refined}}

We now describe these regions in more detail, starting with the deficit regions. As mentioned above, the deficit regions correspond to regions of vanishing kinetic energy of the background flow, together with positive vorticity. The full potential $V$ in 
\eqref{eq:potentials}
is positive in such regions. Few particles have sufficient initial energy to rise on top of such potential hills, hence the deficit in NIW kinetic energy. In Appendix~\ref{app:KEdist} we derive the following refined statistical mechanics prediction for the distribution of NIW kinetic energy, taking into account the narrow distribution of the initial energy $E_0$ of the particles:
\begin{equation}
\overline{|M|_{\text{erg}}^2}(\bx) = \frac{1}{2} \left\{ 1- \text{erf} \left[\frac{\Delta \chi (\bf x) - \gamma |\bnabla \chi|^2 (\bf x)}{\sqrt{2} (\Delta \chi)_{\text{rms}}} \right] \right\} \, , 
\label{eq:predictiondeficit}
\end{equation}
where $(\Delta \chi)_{\text{rms}}$ denotes the rms vorticity of the background flow (rms value of $\Delta \chi$). As illustrated in figure~\ref{fig:comparison_KE_SAR}, this refined prediction accurately captures the location and intensity of the deficit regions. \\

We now turn to the excess regions, which coincide with the vortical cores of the two main anticyclones of the background flow, whose centers are located at $\bx_1=(0.761,0.830)$ and $\bx_2=(0.663,0.107)$. For large $\gamma$, these excess regions occupy an arguably narrow fraction of the domain, and they contain a tiny fraction of the overall NIW kinetic energy (in all the numerical simulations with $\gamma \ge 5$, the overall surplus of energy contained in these two anticyclones is only about 0.5\% of the total energy.).   
The NIW accumulation results from a breaking of ergodicity in the vortical cores of the anticyclones as NIW kinetic energy from the initial condition remains trapped there. Such trapping of NIW kinetic energy can be explained by the existence of localized eigenmodes in the vicinity of the anticyclonic vortex cores, see e.g. the eigenmodes computed by \cite{llewellyn1999near} for an axisymmetric vortex. To test this assumption, we ran a few simulations with a tailored initial condition that features very little NIW kinetic energy in the anticyclones. This initial condition prevents any trapping in the non-ergodic regions, and the regions of excess NIW kinetic energy  are indeed absent from the resulting $\overline{|M|^2}(\bx)$ (not shown). 

To further describe NIW accumulation in the two anticyclones in the large-$\gamma$ limit, we propose an idealized model consisting of classical particles trapped in an axisymmetric vortex. We start from a uniform initial distribution of classical particles in the anticyclone, corresponding to the initial condition $M({\bf x},t=0)=1$. A consequence of axisymmetry is that the subsequent motion of the particles takes place in the radial direction only. In other words, the additional conservation of angular momentum prevents chaotic motion and ergodic statistics for the trapped particles, allowing only for oscillatory radial motion. Early evolution under the YBJ equation leads to $M(\bx,t)\simeq 1-i\gamma\zeta t/2 + {\cal O}(t^2)$, where $\zeta= \Delta \chi$ denotes the vorticity of the background flow (see also \citet{asselin2020refraction}). Denoting as $r$ the radial coordinate, the early-time radial velocity is thus $-i\partial_r M=-\gamma(\partial_r \zeta) t /2$. It points towards the center of the anticyclone ($\partial_r \zeta>0$), indicating accumulation at early time.
To estimate the radius of influence -- or `trapping radius' -- of each anticyclone, we extract the distance between the vortex center and the first point at which $-\bnabla \zeta$ points away from the vortex center. This leads to the trapping radius $R_1 = 0.030$ (resp. $R_2 = 0.029$) for anticyclone 1 (resp. anticyclone 2). 

In Appendix~\ref{app:axisymac} we describe the subsequent radial motion of the trapped classical particles within the (approximately) axisymmetric vortical core. Firstly, we note that the NIW kinetic energy trapped in the two anticyclones represents only a tiny fraction of the total NIW kinetic energy of the system. This fraction is independent of $\gamma$ for large $\gamma$, being equal to the initial kinetic energy contained within the disks of centers $\bx_1$ and $\bx_2$, and radii $R_1$ and $R_2$: $\pi (R_1^2+R_2^2) \simeq 0.005 \ll 1$ (the total NIW kinetic energy being equal to one with our non-dimensionalization). Secondly, we compute the distribution of excess kinetic energy within the two disk-shaped regions, taking into account the vorticity profiles $\zeta_{1,2}(r)$ of the two anticyclones. We obtain
\begin{equation}
    \overline{|M|_{\text{trap},i}^2}(r) = \frac{1}{r} \int_{r}^{R_i} \frac{r_0 \, \mathrm{d}r_0}{\sqrt{\zeta_i(r_0)-\zeta_i(r)} \times \int_{s=0}^{s=r_0} \frac{\mathrm{d}s}{\sqrt{\zeta_i(r_0)-\zeta_i(s)}}} \quad (i=1,2),
    \label{eq:KEwave_trapping}
\end{equation}
with $r \leq R_i$ the distance to the center of anticyclone considered. 
The full prediction for the spatial distribution of NIW kinetic energy is obtained by adding the two excess distributions (\ref{eq:KEwave_trapping}) to the ergodic contribution  (\ref{eq:predictiondeficit}):
\begin{equation}
    \overline{|M|^2}(\bx) = \overline{|M|_{\text{erg}}^2}(\bx) + \overline{|M|_{\text{trap,1}}^2}(|\bx-\bx_1|) + \overline{|M|_{\text{trap,2}}^2}(|\bx-\bx_2|) \, .
    \label{eq:KEwave_SAR_fullpred}
\end{equation}
This prediction is displayed in figure~\ref{fig:comparison_KE_SAR}. Beyond capturing the deficit regions, it reproduces the accumulation observed in the numerical simulations, at least qualitatively. In particular, the trapped kinetic energy gets redistributed approximately as $1/r$ , with $r$ the distance from the vortex center. Hence the narrow regions of intense $\overline{|M|^2}$ inside the two anticyclonic cores in figure~\ref{fig:comparison_KE_SAR}.

\section{Anticyclonic concentration}
\label{sec:concentrationAnticycl}

Interestingly, in neither of the two limits considered in this study does NIW kinetic energy strongly accumulate in anticyclones. 
To illustrate this point, we introduce a measure $\sigma$ of the preferential concentration of NIW kinetic energy in anticyclonic regions, defined as
\begin{align}
    \sigma = \frac{\langle \overline{|M|^2} (-\Delta\chi) \rangle}{\langle \overline{|M|^2} |\Delta\chi| \rangle} \in [-1,1] \, . \label{eq:defsigma}
\end{align}
The quantity $\sigma$ evaluates to $+1$ (resp. $-1$) if $\overline{|M|^2}$ in non-zero in anticyclones (resp. cyclones) only. In the absence of preferential concentration for $\overline{|M|^2}$, we expect $\sigma=0$. We plot $\sigma$ as a function of $\gamma$ in figure~\ref{fig:KEwave_concentration_anticyclones}. 

For weak background flows, the distribution of NIW kinetic energy is almost uniform, with a weak spatial modulation proportional to the streamfunction of the background flow. While vorticity and streamfunction are strongly correlated for monoscale flows consisting of a single wavenumber, these two fields strongly differ from one another in typical 2DNS flows such as the one considered here, see figure~\ref{fig:background_flow}. Nevertheless, as $\gamma$ increases from zero we expect $\sigma$ to increase linearly with $\gamma$, in line with the prediction \eqref{eq:KESDR}. 

For strong background flows, as noted above, some NIW kinetic energy accumulates in the anticyclonic vortex cores, but this accounts for only a small part of the total. Based on the mechanism for non-ergodic trapping in anticyclonic regions discussed in Appendix~\ref{app:axisymac}, we predict a small but finite limiting value for $\sigma$, $\sigma_{\infty} = \lim_{\gamma \to \infty} \sigma=0.046$. For large but finite $\gamma$, the deficits in NIW kinetic energy located in regions of slow cyclonic flow further contribute to increasing $\sigma$.  

With the goal of characterizing the behavior of $\sigma$ for intermediate values of $\gamma$, we derived a uniform  and parameter-free prediction for $\sigma$ vs $\gamma$ under the form of a two-point Padé approximant~\citep{BenderOrszag1999}. The derivation is deferred to Appendix~\ref{app:strong_disperion_2nd_order_intermediate}, where we first extend the weak-$\gamma$ expansion to second order, before combining the $\gamma \ll 1$ and $\gamma \gg 1$ expansions into a single Padé approximant. The resulting approximant is compared to the numerical data in figure~\ref{fig:KEwave_concentration_anticyclones}. It accurately captures the weak preferential concentration of NIW in anticyclones for $\gamma \ll 1$ and $\gamma \gg 1$, while also providing a good approximation to $\sigma$ for intermediate background flow strength $\gamma$.

In summary, both asymptotic behaviors for $\sigma$ indicate that this quantity is largest for intermediate values of $\gamma$. The present study thus provides an explanation for the observations reported by DVB, namely that NIW accumulation in anticyclones is modest for both slow and fast flows (see also~\citet{zhang2023scale}), while it is maximal for flows of intermediate strength. This statement is made quantitative using a two-point Padé approximant matching the weak-flow and strong-flow asymptotic behaviors of $\sigma$.

\begin{figure}
    \centering
    \includegraphics[width=0.5\linewidth]{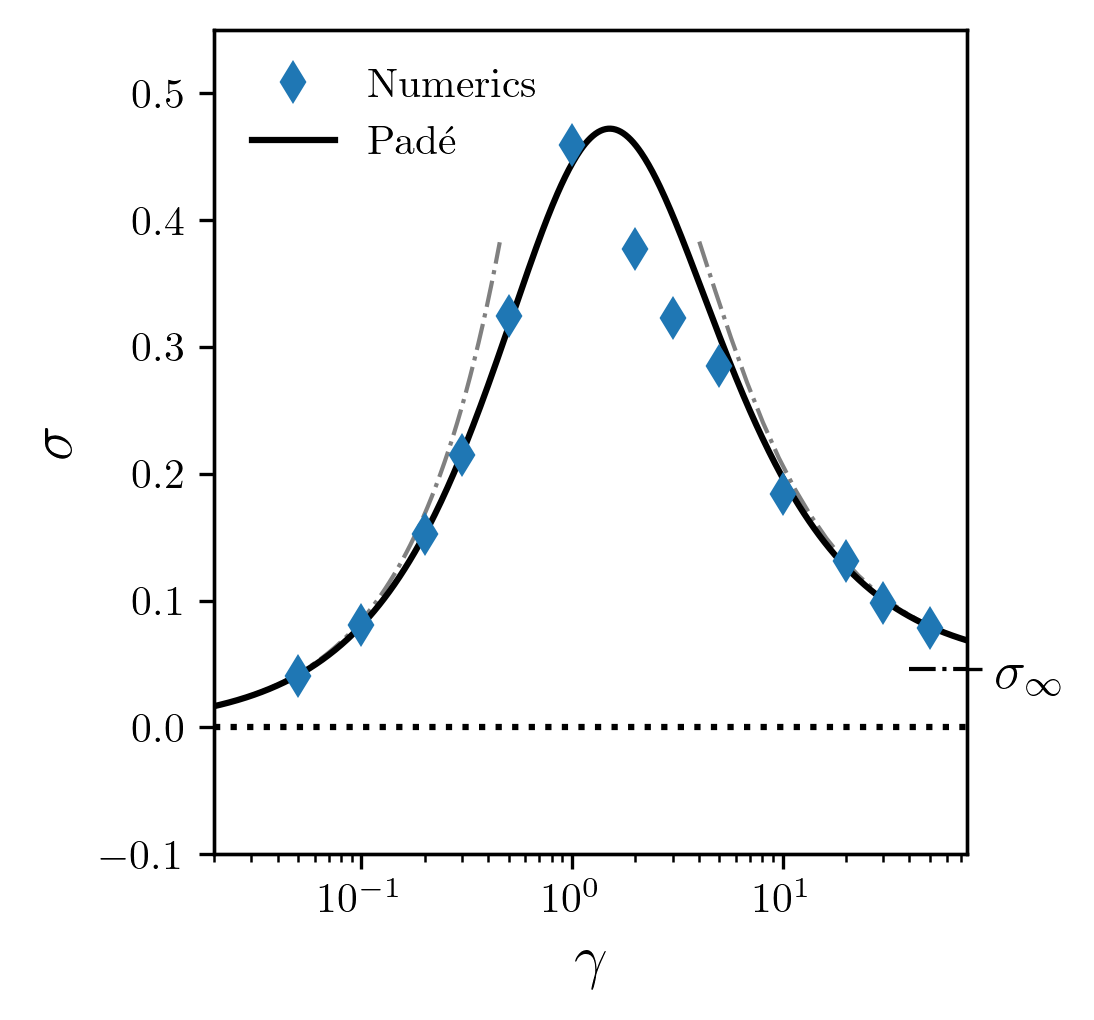}
    \caption{Preferential concentration of NIW kinetic energy in anticyclonic regions as measured by the quantity $\sigma\in[-1,1]$ defined in~(\ref{eq:defsigma}). $\sigma$ is always positive in the simulations (blue diamonds), indicating preferential concentration in anticyclones. Such concentration is maximal for moderate values of $\gamma$ while achieving small values for both small and large $\gamma$. The gray dash-dotted curves show the first-order prediction for slow flows, with $\overline{|M|^2}(\bx)$ given by~\eqref{eq:KESDR}, and the full prediction for fast flows, with $\overline{|M|^2}(\bx)$ given by~\eqref{eq:KEwave_SAR_fullpred}. The solid black curve is the Padé approximant~(\ref{eq:app_Pade}). $\sigma_{\infty}$ is the $\gamma \to \infty$ limiting value of $\sigma$ inferred from trapping in anticyclonic vortex cores. Error bars, estimated from variability across different time-averaging subwindows, are smaller than the symbol size for all simulations.}
    \label{fig:KEwave_concentration_anticyclones}
\end{figure}

\section{Conclusion}

We have revisited the distribution of NIW  over steady background flows within the framework of the YBJ equation, leveraging an analogy with a quantum particle in a steady electromagnetic field. In the limit of weak background flows, we have extended the `strong-dispersion' asymptotic expansion introduced by YBJ to characterize the NIW kinetic energy, potential energy and Stokes drift. We then considered the opposite, `strong-advection' limit of fast background flows. The latter regime corresponds to the classical mechanics limit of the analogous quantum systems. We thus leveraged the statistical mechanics of equilibrium classical systems to predict the spatial distribution of the NIW kinetic energy, potential energy and Stokes drift.

We compared the predictions to numerical simulations of the YBJ equation over a steady background flow consisting of a frozen-in-time velocity field from a simulation of the 2D Navier-Stokes equation. We obtained very good overall agreement with the predictions in both limits, especially for the potential energy and Stokes drift. The strongest departures are obtained for NIW kinetic energy in the strong-advection regime, where trapping within the small anticyclonic vortex cores leads to accumulation. While this accumulation locally leads to large values of the NIW kinetic energy, it represents a negligible fraction of the total NIW kinetic energy. Therefore, most of the NIW kinetic energy is ergodically distributed and agrees with the prediction from statistical mechanics. The velocity field of the present study contrasts with the highly symmetric quadrupolar flow considered by~\citet{zhang2023scale}. A perhaps surprising outcome of the present work is that, in the strong-advection regime, less-organized flows are more easily addressed theoretically than highly-symmetric ones, because the former lead to (predominantly) ergodic statistics.

Of course, the asymptotic limits of large and small flow strength $\gamma$ must remain compatible with the asymptotic expansion leading to the YBJ equation. In Appendix~\ref{app:derivation}, we express the corresponding constraints in terms of the Burger number $\epsilon$, concluding that the YBJ expansion is valid provided $\epsilon \ll \gamma \ll \epsilon^{-1/2}$. Following DVB one may consider the following parameter values for the North Atlantic Ocean, $H_0=100$m, $f=10^{-4}$s$^{-1}$, $g'=2 \times 10^{-3}$m.s$^{-2}$. For a background flow with typical scale $L_\psi=100$km, the resulting Burger number is $\epsilon \simeq 0.002$, suggesting that values of $\gamma$ in the range $[0,01 ; 10]$ are acceptable.

Beyond Ocean values, the asymptotic regimes $\gamma \ll 1 $ and $\gamma \gg 1$ considered in sections~\ref{sec:SDR} and~\ref{sec:classical} correspond to the distinguished limits $\epsilon \ll \gamma \ll 1$ and $1 \ll \gamma \ll \epsilon^{-1/2}$, respectively. 
At first sight, focusing on such asymptotic regimes may seem questionable if one is to make predictions for realistic Ocean flows. Indeed, as discussed by DVB, for $\gamma \gg 1$ the transient time for the system to reach a stationary state (with ergodic statistics) typically exceeds the eddy turnover time of the background flow. Unless the mesoscale background flow is locked to topographic structures on the Ocean floor~\citep{bretherton1976two,bouchet2012statistical,gallet2024two}, one is naturally led to question the steady-flow assumption. However, an indirect reason for studying the $\gamma \gg 1$ asymptotic regime is that, together with the $\gamma \ll 1$ asymptotic expansion, it constrains rather strongly the behavior of the system for more modest, oceanically relevant values of $\gamma$. We have illustrated this phenomenon in section~\ref{sec:concentrationAnticycl} by combining the $\gamma\ll 1$ and the $\gamma \gg 1$ asymptotic expansions into a uniform Padé approximant for the preferential concentration of NIW kinetic energy in anticyclones. The Padé approximant agrees satisfactorily with the numerical data over the entire range of $\gamma$, including the oceanically relevant range $\gamma \sim 1$. 
Of course, it would still be desirable to investigate whether the present approach can be extended to include the temporal dependence and the vertical structure of the background flow, thus narrowing the gap between the present idealized model and true Ocean flows.
In this broader context of 3D time-dependent flows, it remains to be determined what fraction of the NIW energy is fluxed down the anticyclonic drain-pipes as a result of early-time dynamics~\citep{kunze1985near,asselin2020penetration}, potentially inducing wave breaking, what fraction instead equilibrates following the present theoretical arguments, and whether combining large and small $\gamma$ asymptotics into uniform Padé approximants remains a viable approach.

At the methodological level, the main addition of the present work is probably the application of equilibrium statistical mechanics to the ray-tracing system arising in the limit of scale separation. This approach is by no means restricted to NIWs, however, and we hope to report soon on its predictive skill for arbitrary waves in inhomogeneous media.

\backsection[Acknowledgements]{We thank J. Meunier for providing the background flow used in the numerical simulations, and S. Aumaitre for insightful discussions.}

\backsection[Funding]{This research is supported by the European Research Council under grant agreement 101124590.}

\backsection[Declaration of interests]{The authors report no conflict of interest.}

\backsection[Data availability statement]{The scripts to reproduce the numerical simulations used in this study are openly available on Zenodo at https://doi.org/10.5281/zenodo.15632529. Further information can be provided by the authors upon request.}

\backsection[Author ORCIDs]{A. Tlili, https://orcid.org/0009-0000-7901-0022; B. Gallet, https://orcid.org/0000-0002-4366-3889}

\appendix

\section{Derivation of the YBJ equation\label{app:derivation}}

\subsection{Asymptotic expansion}

Only dimensionless quantities are considered throughout this appendix. As in the main text of the article, we omit the tildes to alleviate notations. Consider the complex velocity $\U={u}+i{v}$, whose evolution equation is obtained from the linear combination (\ref{eq:uadim})$+i$(\ref{eq:vadim}):
\begin{align}
\partial_{{\tau}}  \U + i  \U + Ro_{\psi} \left[ {J}({\chi}, \U) + \frac{i}{2} ({\Delta} {\chi})  \U +  \U^* \left(- {\chi}_{ {x} {y}}-\frac{i}{2} {\chi}_{ {y} {y}}+\frac{i}{2} {\chi}_{ {x} {x}} \right) \right] = -\epsilon \left(  {h}_{ {x}} +i  {h}_{ {y}} \right) \, . 
\label{eq:Uadim}
\end{align}

This equation is coupled to the evolution equation for $ {h}$, where $ {u}$ and $ {v}$ are recast in terms of $\U$:
\begin{align}
\partial_{ {\tau}}  {h} + Ro_{\psi}  {J}( {\chi}, {h}) + \partial_{ {x}} \left( \frac{ \U+ \U^*}{2}  \right)+\partial_{ {y}} \left(  \frac{ \U- \U^*}{2 i }  \right) = 0\, .  
\label{eq:hexpand}
\end{align}

Following DVB and YBJ, we fix the ratio $\gamma = Ro_{\psi}/\epsilon = {\cal O}(1)$ and consider a small Burger number $\epsilon \ll 1$ together with a multiple-timescale expansion in $\epsilon$ where all the dimensionless fields are ${\cal O}(1)$ to leading order:

\begin{align}
\U & =\U_0(\bx,\tau,t) + \epsilon \U_1(\bx,\tau,t) + \dots \, , \\
h & = h_0(\bx,\tau,t) + \epsilon h_1(\bx,\tau,t)+ \dots \, . 
\label{eq:scaleh}
\end{align}

The slow time variable $t$ arising in the equations above is defined as $t=\epsilon \tau$.

To leading order, equation~(\ref{eq:Uadim}) reads $\partial_{\tau} \U_0 + i \U_0 = 0$, with solution $\U_0 = M(\bx,t)e^{-i\tau}$. 

To order $\epsilon$, equation~(\ref{eq:Uadim}) reads:
\begin{align}
\nonumber  \partial_{\tau} \U_1 + i \U_1 = & -\partial_t \U_0 - J(\chi,\U_0)-\frac{i}{2} (\Delta \chi) \U_0 - \U_0^* \left[- {\chi}_{ {x} {y}}-\frac{i}{2} {\chi}_{ {y} {y}}+\frac{i}{2} {\chi}_{ {x} {x}} \right] \\
& - \left( \partial_x h_0 + i \partial_y h_0 \right) \, . \label{eq:U1}
\end{align}

The solvability condition requires that there be no resonant term of the form $e^{-i\tau}$ on the rhs. To write this solvability condition, we first need to determine the resonant part of $h_0$, obtained by considering equation~(\ref{eq:hexpand}) at ${\cal O}(1)$ and equal to $(-\frac{i}{2} \partial_x M - \frac{1}{2} \partial_y M ) e^{-i\tau}$ (resonant part only).
Collecting the various contributions proportional to $e^{-i\tau}$ on the rhs of equation~(\ref{eq:U1}) finally yields the solvability condition. This solvability condition is the YBJ equation~(\ref{eq:YBJ}).

\subsection{Range of validity}

A useful starting point to estimate the range of validity of the present study is the work of \cite{thomas2017near}, who derived an extended version of the YBJ equation. As compared to the standard YBJ equation, their equation (3.17) includes additional terms that are subdominant in the YBJ regime $\epsilon \ll 1$ and $\gamma={\cal O}(1)$. However, these new terms can become significant if $\gamma$ is made too large or too small for fixed $\epsilon$. Specifically, equation (3.17) of \cite{thomas2017near} is valid provided $Ro_{\psi}^2 \lesssim \epsilon \lesssim Ro_{\psi}^{1/2}$. When either of these two inequalities is saturated, additional terms need to be incorporated into the YBJ equation. By contrast, when these inequalities are sharply satisfied, $Ro_{\psi}^2 \ll \epsilon \ll Ro_{\psi}^{1/2}$, these additional terms are negligible and the standard YBJ equation holds. We recast this range of validity of the YBJ equation in terms of $\gamma$ and $\epsilon$ using $Ro_\psi=\gamma \epsilon$, finally obtaining the conditions $\epsilon \ll \gamma \ll \epsilon^{-1/2}$.

\section{Stokes drift\label{app:Stokes}}

The Stokes drift of the NIW field is readily defined within the multiple timescale framework of Appendix~\ref{app:derivation}. Denoting as $\overline{ \, \cdot \, }^{\tau}$ an average over the fast time variable $\tau$ of Appendix~\ref{app:derivation} -- or equivalently, over a single inertial period -- the Stokes drift is defined as ${\bf u}_s = \overline{ ({\boldsymbol{\xi}} \cdot \bnabla) {\bf u}}^{\tau}$, where the mean-zero oscillatory displacement field ${\boldsymbol{\xi}}=(\xi_x,\xi_y)$ is defined through $\partial_{\tau} {\boldsymbol{\xi}} = {\bf u}$ (fast-time derivative only). The latter equation can be recast as $\partial_{\tau} (\xi_x + i \xi_y) = {\cal U}$. After inserting the lowest-order complex velocity $\U_0 = M(x,y,t)e^{-i\tau}$ on the rhs, we obtain the lowest-order oscillatory displacements $\xi_x + i \xi_y=i M e^{-i\tau}$, that is:
\begin{align}
\xi_x & = \frac{i M e^{-i\tau} - i M^* e^{i\tau}}{2} \, , \qquad \xi_y = \frac{ M e^{-i\tau} + M^* e^{i\tau}}{2} \, . \label{eq:appcompxi}
\end{align}

The components of the oscillatory velocity field read:
\begin{align}
u & = \frac{M e^{-i\tau} + M^* e^{i\tau}}{2} \, , \qquad v = \frac{ M e^{-i\tau} - M^* e^{i\tau}}{2i} \, , \label{eq:appcompu}
\end{align}

and after substitution of~(\ref{eq:appcompxi}-\ref{eq:appcompu}) into the definition of ${\bf u}_s$ one obtains:
\begin{align}
{\bf u}_s = \frac{1}{4} \left[ i(M \bnabla M^* - M^* \bnabla M) + \bnabla \times (|M|^2 {\bf e}_z) \right]  \, . 
\end{align}
In the main text we are concerned with the long-time average of this quantity, including an average over the slow time $t$. This additional time average leads to equation~(\ref{eq:defus}).

\section{Integration in phase space\label{app:Integrals}}

\subsection{Ergodic computation of the potential energy}

Substituting the probability density~(\ref{eq:Probameasure}) with the prefactor ${\cal C}=1/(2\pi)$ into the integral arising on the rhs of~(\ref{eq:tempintPE}), before changing integration variable to ${\bf K}={\bf p}+ \gamma{\bf U}(\bx)$, leads to:
\begin{align}
\overline{|\bnabla M|^2}(\bx) & = \frac{1}{2\pi} \int_{{\bf p}\in \mathbb{R}^2} p^2  \delta[H(\bx,{\bf p})-E_0] \, \mathrm{d} {\bf p} \, \\
& = \frac{1}{2\pi} \int_{{\bf K}\in \mathbb{R}^2} [{\bf K}-\gamma{\bf U}(\bx)]^2  \delta \left[ \frac{1}{2}K^2  + V(\bx) -E_0  \right] \, \mathrm{d} {\bf K} \, \\
& = \frac{1}{2\pi} \int_{{\bf K}\in \mathbb{R}^2} [K^2-\cancel{2\gamma{\bf K}\cdot {\bf U}(\bx)}+\gamma^2 {\bf U}(\bx)^2] \,  \delta \left[ \frac{K^2}{2}  + V(\bx) -E_0  \right] \, \mathrm{d} {\bf K} \, ,
\end{align}
where the crossed-out term above is odd in ${\bf K}$, thus leading to a vanishing integral. Changing integration variable to $s=K^2/2$, with $\mathrm{d} {\bf K}=2 \pi \mathrm{d}s$, we finally obtain:
\begin{align}
\overline{|\bnabla M|^2}(\bx) & =   \int_0^\infty 2 s \,  \delta \left[s  + V(\bx) -E_0  \right]  \, \mathrm{d} s  + \gamma^2 {\bf U}(\bx)^2  \int_0^\infty    \delta \left[ s  + V(\bx) -E_0  \right]  \, \mathrm{d} s \, \\
& = \underbrace{\left( 2(E_0-V(\bx))+\gamma^2{\bf U}(\bx)^2 \right)}_{\approx 2 \gamma^2 {\bf U}(\bx)^2} \underbrace{{\cal H}(E_0-V(\bx))}_{\approx 1} = 2 \gamma^2  {\bf U}(\bx)^2 \, . 
\end{align}

\subsection{Ergodic computation of the Stokes drift}

To compute the time-averaged Stokes drift, we notice that the following integral vanishes:
\begin{align}
\nonumber \int_{\bx'\in {\cal D}, \, {\bf p}\in \mathbb{R}^2} [{\bf p} + \gamma {\bf U}(\bx)] {\delta(\bx'-\bx)} \, {\cal P}(\bx',{\bf p}) \, \mathrm{d}\bx' \mathrm{d} {\bf p} & = \frac{1}{2 \pi} \int_{ {\bf p}\in \mathbb{R}^2} [{\bf p} + \gamma {\bf U}(\bx)] \, \delta[H(\bx,{\bf k})-E_0]  \, \mathrm{d} {\bf p} \\
& =   \frac{1}{2 \pi} \int_{ {\bf K}\in \mathbb{R}^2} {\bf K} \, \delta \left[\frac{K^2}{2}+V(\bx)-E_0 \right]  \, \mathrm{d} {\bf K} = 0 \, , \label{eq:tempint}
\end{align}
where we have changed integration variable to ${\bf K}={\bf p}+ \gamma {\bf U}(\bx)$ and the last equality stems from the integrand being odd in the components of ${\bf K}$. Rearranging the lhs of (\ref{eq:tempint}) we obtain $\overline{{\bf p} |M|^2}(\bx) = -\gamma{\bf U}(\bx) \overline{|M|^2}(\bx) = -\gamma{\bf U}(\bx)$, where we have substituted the lowest-order prediction $\overline{|M|^2}(\bx)=1$. We finally obtain:
\begin{align}
\overline{{\bf u}_s}(\bx) =- \frac{\gamma}{2} {\bf U}(\bx)  \, . 
\end{align}

\subsection{Kinetic energy: including the distribution of initial energy\label{app:KEdist}}

The approximate prediction~(\ref{eq:predictionM2}) was obtained from~(\ref{eq:tempheavyside}) by assuming that the initial energy $E_0$ vanishes. Weight-averaging the prediction \eqref{eq:tempheavyside} for uniform $E_0$ with the narrow distribution of initial energy ${\cal P}_{E_0}(E_0)$ leads to a refined prediction for the ergodic contribution to $\overline{|M|^2}(\bx)$:
\begin{align}
\overline{|M|_{\text{erg}}^2}(\bx) & = \int_{-\infty}^{\infty}  \, \frac{{\cal H}[E_0-V(\bx)]}{\la {\cal H}[E_0-V(\bx)] \ra}  {\cal P}_{E_0}(E_0)\,  \mathrm{d}E_0 \, , \label{eq:M2weighteddenom}
\end{align}
where the denominator is a normalization factor ensuring the conservation of action among the particles with energy $E_0$. This factor can be computed as the area fraction of the domain that is accessible to particles with initial energy $E_0$, and it is approximately equal to one for all the values of $E_0$ for which ${\cal P}_{E_0}$ has significant weight.
Indeed, the initial energy of a particle is given by half the local ${\cal O}(\gamma)$ vorticity, whereas the potential $V(\bx)$ is dominated by the large negative term $-\gamma^2|\bnabla \chi|^2={\cal O}(\gamma^2)$ almost everywhere. We thus make the approximation $\la {\cal H}[E_0-V(\bx)] \ra \simeq 1$ in the following.
The initial energy being equal to one half the local vorticity $\gamma\zeta$ of the background flow also results in a relation between the pdfs of these two quantities. Denoting the normalized vorticity pdf as ${\cal P}_{\zeta}(\zeta)$, we have ${\cal P}_{E_0}(E_0)\mathrm{d}E_0 = {\cal P}_{\zeta}(\zeta)\mathrm{d}\zeta$ with $E_0 = \frac{1}{2}\gamma \zeta$, which after substitution into~(\ref{eq:M2weighteddenom}) yields:
\begin{align}
\overline{|M|^2_{\text{erg}}}(\bx) = \int_{-\infty}^{\infty}  \, {\cal H}\left[\frac{1}{2}\gamma \zeta-V(\bx)\right]  {\cal P}_{\zeta}(\zeta)\,  \mathrm{d}\zeta = \int_{2V(\bx)/\gamma}^{\infty} {\cal P}_{\zeta}(\zeta) \mathrm{d}\zeta \, .
\label{eq:appintPzeta}
\end{align}

At this stage, various approximations to the distribution ${\cal P}_{\zeta}$ can be considered based on a balance between simplicity and accuracy:
\begin{itemize}
\item The simplest approximation is the one considered in the main text. We note that, for large $\gamma$,  $V(\bx)$ is much more negative than $E_0$ throughout most of the domain. The precise distribution of $E_0$ around zero is then neglected. That is, we substitute the approximation ${\cal P}_{\zeta}=\delta(\zeta)$ into (\ref{eq:appintPzeta}), which leads to (\ref{eq:predictionM2}).
\item A more accurate yet less readable approach consists in substituting the exact vorticity distribution ${\cal P}_{\zeta}$ of the background flow, before computing the integral (\ref{eq:appintPzeta}) numerically. 
\item A trade-off between simplicity and accuracy consists in noting that, for a background flow whose integral scale is small  compared to the domain size, the central limit theorem suggests a normal distribution for the background flow vorticity:
\end{itemize}
\begin{align}
{\cal P}_{\zeta}(\zeta) \simeq \frac{1}{\sqrt{2 \pi} \zeta_{\text{rms}}} e^{-\frac{\zeta^2}{2 \zeta_{\text{rms}}^2}} \, , \label{eq:appnormal}
\end{align}
where $\zeta_{\text{rms}}$ denotes the rms value of $\zeta$.
While the integral scale of our background flow is comparable to the domain size, we adopt the normal distribution and insert~(\ref{eq:appnormal}) into~(\ref{eq:appintPzeta}). Substituting the expression of $V(\bx)$ leads to equation~(\ref{eq:predictiondeficit}).

\section{Initially steady particles in an axisymmetric anticyclone\label{app:axisymac}}

We model the two dominant anticyclones of the background flow as axisymmetric vortices with an increasing vorticity profile up to the radius of influence $R$ introduced in the main text. The particles located at $r<R$ are initially attracted towards the center of the anticyclone and remain trapped. At the level of the YBJ equation, the uniform initial condition for $M$ evolves into an axisymmetric solution $M(r,t)$. For such axisymmetric flow and complex NIW amplitude, the advective term of the YBJ equation vanishes. We are left with a Schrödinger equation describing the radial motion of particles in an axisymmetric potential $\gamma\zeta(r)/2$. The large-$\gamma$, `classical' limit, corresponds to an ensemble of classical particles initially at rest and uniformly distributed. The subsequent motion of the particles is purely radial and conserves mechanical energy. Denoting as $r(t)$ the trajectory of a particle initially located at $r(t=0)=r_0$, the mechanical energy of the particle reads:
\begin{align}
\frac{1}{2}\left( \frac{\mathrm{d}r}{\mathrm{d}t} \right)^2 + \frac{\gamma}{2}\zeta(r) = \frac{\gamma}{2}\zeta(r_0)  \, .
\end{align}
From this equality we extract $\mathrm{d}t$, the infinitesimal time spent by the particle between $r$ and $r+\mathrm{d}r$, as:
\begin{align}
dt= \frac{|\mathrm{d}r|}{\sqrt{\gamma\zeta(r_0)-\gamma\zeta(r)}} \, . \label{eq:appinftime}
\end{align}
Going back to the ensemble of initially uniformly distributed particles, we introduce a distribution $N(r,r_0)$ defined such that $N(r,r_0)\mathrm{d}r\mathrm{d}r_0$ denotes the time-averaged number of particles located between $r$ and $r+\mathrm{d}r$ that were initially located between $r_0$ and $r_0+\mathrm{d}r_0$. Clearly for any given $r_0$ this quantity is proportional to the time~(\ref{eq:appinftime}) spent by a particle between $r$ and $r+\mathrm{d}r$, so that:
\begin{align}
N(r,r_0)= \frac{{\cal C}(r_0)}{\sqrt{\zeta(r_0)-\zeta(r)}} \, ,
\end{align}
where the function ${\cal C}(r_0)$ is determined by the constraint:
\begin{align}
\left( \int_{r=0}^{r=r_0} N(r,r_0) \, \mathrm{d}r \right) \mathrm{d}r_0 = 2 \pi r_0 \, \mathrm{d}r_0 \, . \label{eq:appconstraintnorm}
\end{align}
On the left-hand side of~(\ref{eq:appconstraintnorm}) is the total number of particles that were initially released between $r_0$ and $r_0+\mathrm{d}r_0$. For a uniform initial condition $M(r,t=0)=1$, this number of particles is also equal to $|M(r_0,t=0)|^2 2 \pi r_0 \, \mathrm{d}r_0 = 2 \pi r_0 \, \mathrm{d}r_0 $, hence the equality~(\ref{eq:appconstraintnorm}). ${\cal C}(r_0)$ is easily obtained from~(\ref{eq:appconstraintnorm}), leading to:
\begin{align}
N(r,r_0)= \frac{2\pi r_0}{\int_{s=0}^{s=r_0} \frac{\mathrm{d}s}{\sqrt{\zeta(r_0)-\zeta(s)}}   }   \times   \frac{1}{\sqrt{\zeta(r_0)-\zeta(r)}} \, . \label{eq:appdistr}
\end{align}
On the one hand, the time-averaged number of particles located between $r$ and $r+\mathrm{d}r$ is given by $\left(\int_{r_0=r}^{r_0=R} N(r,r_0) \mathrm{d}r_0 \right) \mathrm{d}r$. On the other hand, it is simply equal to $\overline{|M|^2}(r) 2 \pi r \mathrm{d}r$. Equating the two expressions and inserting (\ref{eq:appdistr}) yields the desired expression for the time-averaged excess kinetic energy:
\begin{align}
\overline{|M|^2}(r) = \frac{1}{r} \int_{r_0=r}^{r_0=R} \frac{r_0}{\int_{s=0}^{s=r_0} \frac{\mathrm{d}s}{\sqrt{\zeta(r_0)-\zeta(s)}} } \times \frac{1}{\sqrt{\zeta(r_0)-\zeta(r)}} \, \mathrm{d}r_0 \, . \label{eq:predictedexcess}
\end{align}
One can check that $\int_0^R \overline{|M|^2}(r) 2 \pi r \mathrm{d}r=\pi R^2$, corresponding to the initial NIW kinetic energy contained inside the non-ergodic trapping region. Expression~(\ref{eq:predictedexcess}) corresponds to the excess kinetic energy due to trapping in non-ergodic regions. This expression behaves as $1/r$ for low $r$, and it vanishes for $r=R$. Of course, beyond this strictly classical computation we expect the $1/r$ divergence to be regularized by quantum effects for small enough $r$, when the latter is comparable to the de Broglie wavelength of the particles. This regularization does not impact the total wave action contained in the trapping region, nor does it affect the resulting value of $\sigma$, and therefore we omit it for brevity.

The full prediction~(\ref{eq:KEwave_SAR_fullpred}) for $\overline{|M|^2}(\bx)$ corresponds to the refined ergodic prediction~(\ref{eq:predictiondeficit}), to which we add the non-ergodic correction~(\ref{eq:predictedexcess}) around the two main anticyclones. Specifically, the vorticity profiles of the two anticyclones are reasonably well fit by the profiles $\zeta_{1;2}(r)=-Z_{1;2} \exp\left( -r^2/(a_{1;2}+b_{1;2} r) \right)$ up to their respective influence radii, with the parameter values $(Z_1,a_1,b_1)=(49, 0.0001, 0.017)$ for anticyclone 1 and $(Z_2,a_2,b_2)=(45, 0.0005, 0.008)$ for anticyclone 2. We compute the excess distribution~(\ref{eq:predictedexcess}) around each anticyclone using this theoretical profile and the parameters $(Z,a,b,R)$ corresponding to each anticyclone. 

For non-axisymmetric anticyclones the angular momentum of individual particles is no longer conserved. In principle, this can lead to chaotic trajectories. Nevertheless, for nearly axisymmetric flows we expect some KAM tori to persist, preserving particle trapping and non-ergodic motion within the anticyclone. We checked this by numerically integrating the ray-tracing equations~\eqref{eq:Hequations} inside an anticyclone with elliptic Gaussian streamfunction. For low to moderate ellipticity, most particles initially located within the disk $r<R$ remain trapped, in  agreement with our axisymmetric description.
As for the axisymmetric case, an initially uniform distribution of particles inside this disk becomes concentrated in a narrow region at the center of the elliptical anticyclone over long-time average. In other words, the prediction~\eqref{eq:predictedexcess} continues to provide qualitative insight for near-axisymmetric anticyclones.

\section{Higher-order expansion and Padé approximant\label{app:strong_disperion_2nd_order_intermediate}}

\subsection{Strong-dispersion expansion to second order in $\gamma$}

With the goal of determining the time-averaged distribution of NIW kinetic energy up to second order in $\gamma \ll 1$, we recast the YBJ equation under the form:
\begin{equation}
    \partial_t M - \frac{i}{2}\Delta M = -\gamma \left( J(\chi,M) + \frac{i}{2}(\Delta\chi)M \right) \, ,
    \label{eq:YBJ_SDR}
\end{equation}
before introducing a multiple-timescale expansion of $M$ as:
\begin{equation}
M({\bf x},t)=M_0({\bf x},t,T_2)+\gamma M_1({\bf x},t,T_2)+\gamma^2 M_2({\bf x},t,T_2)+{\cal O}(\gamma^3) \, , \label{eq:expansiongamma2}
\end{equation}
where $T_2=\gamma^2 t$. Throughout this appendix we refer to $t$ as the fast time variable. It should not be confused with the faster time variable $\tau$ associated with inertial motion, which does not appear in the present appendix.

To ${\cal O}(1)$, equation (\ref{eq:YBJ_SDR}) yields $\partial_t M_0 - \frac{i}{2}\Delta M_0=0$, with solution $M_0={\cal M}_0(T_2)$ (in line with the main text, we denote by a cursive ${\cal M}$ the $x$-independent contributions).

To ${\cal O}(\gamma)$, equation (\ref{eq:YBJ_SDR}) yields:
\begin{equation}
    \partial_t M_1 - \frac{i}{2}\Delta M_1 = - \frac{i}{2}(\Delta\chi){\cal M}_0 \, ,
\end{equation}
with solution:
\begin{equation}
M_1 = {\cal M}_0(T_2) \left[ \chi({\bf x}) + \tilde{m}_1({\bf x},t) \right] + {\cal M}_1(T_2) \, ,
\end{equation}
where $\tilde{m}_1$ is given by the rhs of equation~(\ref{eq:exprmtilde}). One can show that the dynamics of $\mathcal{M}_1(T_2)$ has no influence on the prediction for $\overline{|M|^2}$ at the desired order. In the interest of brevity, we thus set $\mathcal{M}_1(T_2)=0$ in the following.

To ${\cal O}(\gamma^2)$, equation (\ref{eq:YBJ_SDR}) yields:
\begin{equation}
    \partial_t M_2 - \frac{i}{2}\Delta M_2 =-\partial_{T_2} {\cal M}_0 - {\cal M}_0 J(\chi,\tilde{m}_1) - \frac{i {\cal M}_0}{2}(\Delta\chi) (\chi + \tilde{m}_1) \, . \label{eq:ordergamma2}
\end{equation}
Averaging over space and fast time $t$ yields the solvability condition
\begin{equation}
0 = -\partial_{T_2} {\cal M}_0 - i \la (\Delta \chi) \chi \ra {\cal M}_0/2 \, , \label{eq:SCordergamma2}
\end{equation}
whose solution satisfying the initial condition $M({\bf x},0)=1$ is ${\cal M}_0(T_2)=\exp \left[- i \, T_2 \la (\Delta \chi) \chi \ra /2 \right]$. Subtracting (\ref{eq:SCordergamma2}) from (\ref{eq:ordergamma2}) before averaging over the fast time variable $t$ (denoting as $\overline{\, \cdot \, }^t$ this average) yields, after multiplication by $2i$: 
\begin{equation}
 \Delta \overline{M_2}^{t} = {\cal M}_0 (\chi \Delta \chi-\la \chi \Delta \chi \ra) \, ,
\end{equation}
whose solution for the fast-time average of $M_2$ is:
\begin{equation}
\overline{M_2}^{t} = {\cal M}_0  \Delta^{-1} \left\{ \chi \Delta \chi-\la \chi \Delta \chi \ra \right\} + \overline{{\cal M}_2}^{t}(T_2) \, ,
\end{equation}
where $\overline{{\cal M}_2}^{t}(T_2)$ is unknown at this stage. Time-averaging the squared modulus of expansion (\ref{eq:expansiongamma2}) gives, using $\overline{\tilde{m}_1}^{t}=0$:
\begin{equation}
\overline{|M|^2}^{t}({\bf x})=1+2\gamma \chi + \gamma^2 \left( \chi^2+ \overline{|\tilde{m}_1|^2}^{t} + 2 \Delta^{-1} \left\{ \chi \Delta \chi-\la \chi \Delta \chi \ra \right\} + \overline{{\cal M}_2+{\cal M}_2^*}^{t} \right) + {\cal O}(\gamma^3)\, .  \label{eq:modulusexp}
\end{equation}
The space average of this quantity equals one as a result of action conservation, which yields $\overline{{\cal M}_2+{\cal M}_2^*}^{t}=-\la \chi^2 \ra - \la \overline{|\tilde{m}_1|^2}^{t} \ra = -2 \la \chi^2 \ra$. Inserting this expression for $\overline{{\cal M}_2+{\cal M}_2^*}^{t}$ back into~(\ref{eq:modulusexp}) finally yields the sought prediction for the time-averaged NIW kinetic energy: 
\begin{equation}
\overline{|M|^2}^{t}({\bf x})=1+2\gamma \chi + \gamma^2 \left( \chi^2+ \overline{|\tilde{m}_1|^2}^{t} + 2 \Delta^{-1} \left\{ \chi \Delta \chi-\la \chi \Delta \chi \ra \right\} -2 \la \chi^2 \ra \right) + {\cal O}(\gamma^3)\, , \label{eq:expmodMorder2}
\end{equation}
where:
\begin{equation}
    \overline{|\tilde{m}_1|^2}^{t}(\bx) = \sum_{\bk,\boldsymbol{\ell}} \delta_{k^2,\ell^2} \hat{\chi}_{\bk} \hat{\chi}_{\boldsymbol{\ell}}^* e^{i(\bk-\boldsymbol{\ell})\boldsymbol{\cdot x}} \, .
\end{equation}

A limitation of the present expansion is that $\tilde{m}_1$ induces resonant terms on the rhs of equation (\ref{eq:ordergamma2}). These can be dealt with by allowing the spatial Fourier amplitudes of $\tilde{m}_1$ to evolve with the intermediate time variable $T_1=\epsilon t$, but the resulting evolution equations are quite cumbersome. When this dependence is forgotten about, equation~(\ref{eq:expmodMorder2}) remains valid as an average over the fast time variable $t$, and up to a time horizon of order $1/\gamma$. This prediction seems satisfactory when compared to long-time averages of the numerical results, and therefore we omit the $T_1$ and $T_2$ dependence for simplicity.

\subsection{A Padé approximant for $\sigma$\label{app:Pade}}

On the one hand, inserting expression~(\ref{eq:expmodMorder2}) into~(\ref{eq:defsigma}) and performing a few integrations by parts yields a low-$\gamma$ expansion of $\sigma$ of the form $\sigma = \sigma_1 \gamma + \sigma_2 \gamma^2 + {\cal O}(\gamma^3)$, where:
\begin{equation}
\sigma_1 = 2 \frac{\langle |\bnabla \chi|^2 \rangle }{\langle |\Delta \chi| \rangle} \, , \qquad \sigma_2 = -  \frac{3 \la \chi^2 \Delta\chi  \ra + \la \overline{|\tilde{m}_1|^2} \Delta\chi \ra}{\langle |\Delta\chi| \rangle} - 4\frac{ \langle \chi |\Delta\chi| \rangle \langle |\bnabla \chi|^2 \rangle }{\langle |\Delta\chi| \rangle^2} \, .
\end{equation}
On the other hand, inserting the full high-$\gamma$ prediction~(\ref{eq:KEwave_SAR_fullpred}) into~(\ref{eq:defsigma}) yields a high-$\gamma$ expansion of $\sigma$ of the form $\sigma=\sigma_\infty+\sigma_{-1}/\gamma + o(\gamma^{-1})$. An explicit expression for $\sigma_{-1}$ in terms of $\chi$ appears cumbersome and not particularly insightful. Instead, we extract the theoretical value of $\sigma_{-1}$ numerically by inserting the full theoretical prediction~\eqref{eq:KEwave_SAR_fullpred} evaluated for a few large values of $\gamma$ into the definition~(\ref{eq:defsigma}) of $\sigma$.

With the coefficients $\sigma_1$, $\sigma_2$, $\sigma_{-1}$ and $\sigma_\infty$ at hand, we form a two-point Padé approximant that matches both the low-$\gamma$ expansion and the high-$\gamma$ expansion~\citep{BenderOrszag1999}, under the form:
\begin{equation}
    \sigma = \frac{\alpha_1 \gamma + \alpha_2 \gamma^2}{1 + \beta_1 \gamma + \beta_2 \gamma^2} \, ,
    \label{eq:app_Pade}
\end{equation}
where:
\begin{align}
\alpha_1  = \sigma_1 \, , \qquad  \alpha_2  = \sigma_{\infty}\frac{\sigma_1^2 + \sigma_{\infty}\sigma_2}{\sigma_{\infty}^2 +\sigma_1 \sigma_{-1}} \, , \qquad  \beta_1 = \frac{\sigma_1\sigma_{\infty}-\sigma_2\sigma_{-1}}{\sigma_{\infty}^2 +\sigma_1 \sigma_{-1}} \, , \qquad     \beta_2  = \frac{\sigma_1^2+\sigma_2\sigma_{\infty}}{\sigma_{\infty}^2 +\sigma_1 \sigma_{-1}} \, .
\end{align}

\section{Details on numerical simulations}
\label{app:numerics}

As described in Section~\ref{sec:numerics}, the computations are performed using a pseudo-spectral formulation on a GPU with de-aliasing, constant timestep and no damping term. For most values of $\gamma$, time integration employs a standard fourth-order Runge–Kutta (RK4) scheme. For low $\gamma$ values ($\gamma \le 0.1$), the stiff term $\frac{i}{2}\Delta M$ in \eqref{eq:YBJ} is integrated exactly using the Integrating Factor method \citep{Lawson1967} combined with an RK4 scheme, to ensure numerical stability. The constant time step $dt$, spectral resolution, and time-averaging window $[t_{\min}, \,t_{\max}]$ used to compute $\overline{|M|^2}$, $\overline{|\bnabla M|^2}$ and $\overline{{\bf u}_s}$ are summarized in table~\ref{tab:numerics}.

\begin{table}
  \centering
  \begin{tabular}{|cccc|}
    \toprule
    $\gamma$ & Time-step $dt$ & Resolution & $[t_{\min}, \,t_{\max}]$ \\
    \midrule
    $0.05$ & $2.0\times10^{-4}$ & $1024\times1024$ & $[20, \, 100]$ \\  
    $0.1$ & $2.0\times10^{-4}$ & $1024\times1024$ & $[40, \, 100]$ \\  
    $0.2$ & $1.0\times10^{-5}$ & $1024\times1024$ & $[50, \, 100]$ \\
    $0.3$ & $6.7\times10^{-6}$ & $1024\times1024$ & $[33, \, 67]$ \\ 
    $0.5$ & $4.0\times10^{-6}$ & $1024\times1024$ & $[20, \, 40]$ \\ 
    $1$ & $2.0\times10^{-5}$ & $1024\times1024$ & $[10, \, 30]$ \\ 
    $2$ & $1.0\times10^{-5}$ & $1024\times1024$ & $[5.0, \, 15]$ \\ 
    $3$ & $1.7\times10^{-5}$ & $1024\times1024$ & $[13, \, 33]$ \\ 
    $5$ & $1.0\times10^{-5}$ & $1024\times1024$ & $[4.0, \, 10]$ \\ 
    $10$ & $5.0\times10^{-6}$ & $1024\times1024$ & $[1.0, \, 5.0]$ \\ 
    $20$ & $1.3\times10^{-6}$ & $2048\times2048$ & $[1.0, \, 2.5]$ \\ 
    $30$ & $6.7\times10^{-7}$ & $2048\times2048$ & $[1.0, \, 3.3]$ \\ 
    $50$ & $1.0\times10^{-6}$ & $2048\times2048$ & $[1.0, \, 4.0]$ \\ 
    \bottomrule
  \end{tabular}
    \caption{Parameter values employed in the numerical simulations. The time-averaging window is denoted by $[t_{\min}, \,t_{\max}]$. \label{tab:numerics}}

\end{table}

\bibliographystyle{jfm}
\bibliography{jfm}

\end{document}